\documentclass[apj,onecolumn,tighten]{emulateapj}
\usepackage{mathrsfs}
\usepackage{amsmath}
\begin{document}

\title{Numerical Simulations of Optically Thick Accretion onto a Black Hole - II. Rotating Flow}
\shorttitle{Optically Thick Black Hole Accretion}
\shortauthors{Fragile, et al.}
\author{P. Chris Fragile\footnotemark[1], Ally Olejar}
\affil{Department of Physics \& Astronomy, College of Charleston, Charleston, SC 29424, USA}
\email{fragilep@cofc.edu}
\and
\author{Peter Anninos}
\affil{Lawrence Livermore National Laboratory, P.O. Box 808, Livermore, CA 94550, USA}

\begin{abstract}
In this paper we report on recent upgrades to our general relativistic radiation magnetohydrodynamics code, {\em Cosmos++}, including the development of a new primitive inversion scheme and a hybrid implicit-explicit solver with a more general closure relation for the radiation equations.  The new hybrid solver helps stabilize the treatment of the radiation source terms, while the new closure allows for a much broader range of optical depths to be considered.  These changes allow us to expand by orders of magnitude the range of temperatures, opacities, and mass accretion rates, and move a step closer toward our goal of performing global simulations of radiation-pressure-dominated black hole accretion disks.  In this work we test and validate the new method against an array of problems.  We also demonstrate its ability to handle super-Eddington, quasi-spherical accretion.  Even with just a single proof-of-principle simulation, we already see tantalizing hints of the interesting phenomenology associated with the coupling of radiation and gas in super-Eddington accretion flows.
\end{abstract}
\maketitle

\footnotetext[1]{KITP Visiting Scholar, Kavli Institute for Theoretical Physics, Santa Barbara, CA.}

\section{Introduction}
\label{sec:intro}

In recent years, one of the primary areas of active numerical code development within astrophysics has been in multi-dimensional radiation hydrodynamics \citep[e.g.][]{farris08,muller10,shibata11,zanotti11,fragile12,jiang12,lentz12,sadowski13}.  Radiation plays a critical role in many astrophysical settings, including the interiors of stars, some accretion flows, and most explosive events.  However, multi-dimensional radiation hydrodynamics is very challenging computationally, owing to the large number of degrees of freedom and wide range of temporal and spatial scales present.  Nevertheless, there are many interesting phenomena associated with the nonlinear interaction of radiation and gas that require a numerical treatment for a more complete understanding.  One such application is the study of radiation-pressure dominated accretion flows onto black holes, which is the ultimate goal of the present work.

In working toward this goal, our first step was to modify our relativistic MHD code {\em Cosmos++} \citep{anninos05} to treat radiative processes in black hole accretion disks in the optically-thin limit \citep{fragile09}.  Optically thin treatments are the simplest to implement, as the radiation only enters the hydrodynamic equations as a cooling term.  Such treatments are appropriate for very low accretion rate systems, such as Sgr A*, where we recently applied this technique \citep{dibi12,drappeau13}.  

Then, following the work of \citet{farris08}, we generalized {\em Cosmos++} even further by implementing a two-moment closure formalism for general relativistic radiation hydrodynamics appropriate for the optically-thick limit \citep{fragile12}.  The restriction of that approach to optically-thick flows means that it is only applicable to problems with high degrees of symmetry, e.g. Bondi \citep{fragile12} or Bondi-Hoyle \citep{zanotti11} accretion, and at high accretion rates.  

The obvious next step is to develop a method that functions across a wide range of optical depths, which would allow intermediate mass accretion rates and less symmetric problems to be considered.  Doing so requires two advances beyond the method of \citet{farris08}:  first, a more general closure relation for the radiation moments must be implemented; second, the radiation equations must be solved in an implicit, or at least semi-implicit, way.  In a semi-implicit scheme, which is also referred to as a hybrid explicit-implicit scheme, an implicit step is used to solve the radiation source terms, while an explicit step is used for the rest of the update \citep[c.f.][]{turner01,roedig12}.  The advantage is that an implicit update is expected to be stable and avoids the ``stiffness'' problem associated with the radiation source term, especially when the gas and radiation are close to thermal equilibrium.  The advantage of a semi-implicit scheme, as opposed to fully implicit one, is first that it can be more easily integrated into existing explicit MHD codes.  Second, when only the source term is being treated implicitly, it can be calculated locally, and therefore the method does not require a parallel matrix solve across the entire problem domain, as would a fully implicit scheme. 

In this paper, we present a new way to perform the semi-implicit radiation source calculation.  It involves taking 1st order Taylor expansions of the conserved variables and radiation source terms.  This results in a 9-dimensional matrix equation that, when inverted, returns the updated primitive variables at the new time, completing the update.  This scheme is, effectively, an extension of the 5-dimensional primitive inversion scheme introduced in \citet{noble06}.

This paper also presents a new application of our code to quasi-spherical accretion onto a black hole.  For this simulation, we start from initial conditions similar to the Bondi accretion problem that we have considered before, but we imbue the gas with a small amount of angular momentum, thus breaking the spherical symmetry.  The angular momentum is not enough for significant amounts of the gas to circularize; thus, angular momentum transport, such as from the magneto-rotational instability (MRI), is not required for the gas to accrete onto the black hole.  Nevertheless, the angular momentum is enough for the resulting flow to develop a disk-like structure, as well as a latitude-dependent optical depth and flux.  Thinking about it from an observer's perspective, this would result in the source having a latitude-dependent inferred luminosity.  In this work we present a case with an accretion rate of ten times the Eddington rate, and find that the radiative flux varies by about 4\% from along the symmetry axis to the midplane.

Section \ref{sec:method} describes our method, with particular emphasis on the new closure relation and semi-implicit method for solving the radiation source terms.  In Section \ref{sec:tests}, we report on a series of test problems meant to validate our code.  In Section \ref{sec:proga}, we arrive at the novel new result of this paper -- two-dimensional, quasi-spherical accretion onto a black hole, including radiation. We conclude in Section \ref{sec:conclusion}. Most of the equations in this work are written in units where $GM = c = 1$, although in a few places we leave in factors of $c$ for clarity.  These are also the only unit restrictions in the code; fluid and radiation variables can be evolved in otherwise arbitrary units.  Opacities and temperatures, however, are normally tabulated in cgs units.

\section{Numerical Method}
\label{sec:method}

Since most of our numerical method remains the same as was presented in \citet{fragile12}, we give only an abbreviated presentation here. We also restrict discussion to the radiation-hydrodynamics equations, ignoring magnetic fields which do not impact the new method beyond what has already been discussed in our previous paper.

\subsection{Primitive Variables}
\label{sec:primitives}

The principle change from \citet{fragile12} is that we now treat the radiation in its own rest frame (defined as the frame in which the radiation flux vanishes), rather than in the fluid rest frame.  This approach was introduced in the recent paper \citet{sadowski13}.  Its advantages will become apparent in a moment.

To keep the notation distinct between radiation variables in the different frames, we introduce two new variables, $E_R$ and $u_R^i$, representing the radiation energy density in the radiation rest frame, and the spatial components of the radiation rest frame 4-velocity, respectively.  Formally, the radiation rest frame transport velocity, $\vert V_R \vert =\sqrt{(u_R)_i u_R^i}/u_R^t$, can not go to $c$ in our method.  In the same way that we must place a limiter on the fluid transport velocity (or equivalently its boost factor), we also limit the radiation velocity. Thus, there is always a frame in which we assume the radiation is at rest and isotropic, as required by the $\bf{M}_1$ closure. By extension, this means we do not formally reach the free-streaming limit, though we can come arbitrarily close under ideal circumstances.

The new variables can, of course, be related to our original ones, $E$ and $F^i$, representing the radiation energy density and radiation flux in the fluid frame, respectively.  To do so, we can compare the radiation stress energy defined in terms of the two sets of variables.  In terms of the variables used in our previous paper, the covariant radiation stress energy tensor is
\begin{equation}
R^{\alpha \beta} = E u^\alpha u^\beta + F^\alpha u^\beta + F^\beta u^\alpha + P_\mathrm{rad} h^{\alpha \beta} ~,
\label{eq:radstress_old}
\end{equation}
where $h^{\alpha \beta} = g^{\alpha \beta} + u^\alpha u^\beta$ is the projection tensor, $g^{\alpha\beta}$ is the spacetime 4-metric, $u^\alpha$ is the fluid rest frame 4-velocity, and the flux satisfies the normalization $F^\alpha u_\alpha = 0$.  In order to close this expression, we must define the radiation pressure in terms of $E$ (and possibly $F^i$).  In our previous work, we utilized the so-called Eddington approximation, $P_\mathrm{rad} = E/3$, which assumes the radiation pressure is isotropic in the fluid frame.  In terms of our new variables, the radiation stress tensor becomes \citep{sadowski13}
\begin{equation}
R^{\alpha \beta} = \frac{4}{3} E_R u^{\alpha}_R u^{\beta}_R + \frac{1}{3} E_R g^{\alpha \beta}~.
\label{eq:radstress}
\end{equation}
Note that the radiation pressure does not appear explicitly in this expression.  That is because it represents the covariant formulation of the $\bf{M}_1$ closure scheme \citep{levermore84,sadowski13}, which assumes that the radiation is isotropic in the radiation rest frame. 

If we wish to convert from the old variables to the new ones, we can easily solve the following two equations for $E_R$ and $u_R^t$ given the radiation stress tensor components $R^{t\nu}$ defined using equation (\ref{eq:radstress_old}) \citep{sadowski13}:
\begin{eqnarray}
g_{\mu\nu} R^{t \mu} R^{t \nu} & = & -\frac{8}{9} E_R^2 (u^t_R)^2 + \frac{1}{9} E_R^2 g^{tt} \\
R^{tt} & = & \frac{4}{3} E_R (u^t_R)^2 + \frac{1}{3} E_R g^{tt} ~.
\label{eq:rtt}
\end{eqnarray}
The time components of equation (\ref{eq:radstress}) can be used to find the remaining spatial components, $u_R^i$.

That defines the new primitive radiation variables.  However, we also changed our code to use a different form of the fluid and radiation velocities.  Now, during the evolution, we use the fluid and radiation 4-velocities projected into the space of the normal observer, i.e. an observer with 4-velocity $(-\alpha, 0, 0, 0)$.  The contravariant time component of the 4-velocity vanishes under this projection, while the spatial components become
\begin{equation}
\widetilde{u}^i =  u^i - u^t \frac{g^{ti}}{g^{tt}} = u^i + \frac{\gamma}{\alpha} \beta^i ~,
\end{equation}
\begin{equation}
\widetilde{u}^i_R = u^i_R  - u^t_R \frac{g^{ti}}{g^{tt}} ~,
\end{equation}
where $\alpha^2 = -1/g^{tt}$ is the square of the lapse, $\beta^i = \alpha^2 g^{ti}$ is the shift vector, $\gamma = \sqrt{1 + g_{ij} \widetilde{u}^i \widetilde{u}^j}$ is the Lorentz factor of the flow as measured by the normal observer, and $u^t = \gamma/\alpha$.  Compared with the transport velocity, $V^i = u^i/u^t$, which is restricted to the numerical range $-c \le V^i \le c$, $\widetilde{u}^i$ has the advantage that it has no physical restriction on its range, going from $-\infty$ to $+\infty$.  The other possible choice for a primitive velocity variable, $u^i$, has the problem that the resulting expression for $u^t$ can be ambiguous as to a sign.  With these changes, the set of primitive variables used in this work is:
\begin{equation}
\mathbf{P} = \left( \begin{array}{c} \rho \\ \epsilon \\ \widetilde{u}^i \\ E_R \\ \widetilde{u}^i_R \end{array} \right) ~,
\end{equation}
where $\rho$ is the rest mass density and $\epsilon$ is the specific internal energy, both measured in the fluid rest frame. This represents a change from $\mathbf{P}=(\rho,~\rho\epsilon, ~V^i, ~E, ~F^i)$ used in our previous work. The switch from $\rho\epsilon$ to $\epsilon$ simplifies the Taylor expansions used for solving the source term and primitive inversion in Section \ref{sec:source}; the switch from $V^i$ to $\widetilde{u}^i$ adds robustness and stability in certain cases; and finally, the switch from $(E,~F^i)$ to $(E_R, ~\widetilde{u}^i_R)$ moves us from the Eddington closure to the $\bf{M}_1$ closure.

\subsection{Evolution Equations}

\subsubsection{Decoupled Radiation}

This new radiation method is formally independent of optical depth; the M1 closure itself does not place any restriction on the optical depth in any frame.  In fact, there are some cases where we may want to treat the radiation variables entirely independently of whatever the background hydrodynamic flow may be doing.  Our beam-of-light test in Section \ref{sec:beam} is one such example.  In this case, it is sufficient to solve the decoupled radiation stress energy equation
\begin{equation}
\left(R^\beta_\alpha\right)_{;\beta} = 0 ~.
\end{equation}
This can be written as the following set of conservation laws:
\begin{eqnarray}
 \partial_t {\cal R} + \partial_i \left(\sqrt{-g}~R^i_t\right) &=&
      \sqrt{-g}~R^\alpha_\beta~\Gamma^\beta_{t \alpha} ~, \\
 \partial_t {\cal R}_j + \partial_i \left(\sqrt{-g}~R^i_j\right) &=&
      \sqrt{-g}~R^\alpha_\beta~\Gamma^\beta_{j \alpha} ~,
\end{eqnarray}
where ${\cal R} = \sqrt{-g}R^t_t$ is the conserved radiation energy density, ${\cal R}_j = \sqrt{-g} R^t_j$ is the conserved radiation momentum density, $g$ is the 4-metric determinant, and $\Gamma^\beta_{\alpha\gamma}$ is the geometric connection coefficients of the metric.  The form of these conservation laws is identical to the form of the fluid energy and momentum conservation laws already solved in {\em Cosmos++}, and the same techniques can be used, specifically the high-resolution shock-capturing (HRSC) scheme, described in \citet{fragile12}.  In this decoupled case, the recovery of the primitive hydrodynamic variables proceeds in the normal way (available options within {\em Cosmos++} are described in \citet{fragile12}), while the primitive radiation variables are recovered from equations (\ref{eq:radstress})--(\ref{eq:rtt}), using the procedure described in Section \ref{sec:primitives}, with the radiation stress energy components coming from the updated conserved variables.

\subsubsection{Coupled Radiation Hydrodynamics}

For more interesting problems, where the radiation and hydrodynamics are coupled, we aim to solve the following set of conservation equations for mass
\begin{equation}
\left(\rho u^\beta\right)_{;\beta} = 0 ~,
\label{eqn:continuity}
\end{equation}
fluid stress-energy
\begin{equation}
\left( T^\beta_\alpha\right)_{;\beta} = G_\alpha ~,
\label{eqn:Tfluid}
\end{equation}
and radiation stress-energy 
\begin{equation}
\left( R^\beta_\alpha\right)_{;\beta} = -G_\alpha ~.
\label{eqn:Trad}
\end{equation}
As usual, the fluid stress-energy tensor is
\begin{equation}
T^{\alpha \beta} = (\rho + \rho \epsilon + P_\mathrm{gas}) u^\alpha u^\beta + P_\mathrm{gas} g^{\alpha \beta} ~,
\end{equation}
where $P_\mathrm{gas}$ is the gas pressure.  In this work we are ignoring magnetic fields, though they can easily be included as described in \citet{fragile12}.  The coupling of the fluid and radiation equations occurs through the radiation 4-force density, $G^\mu$, which can conveniently be written in the form \citep{sadowski14}
\begin{equation}
G^\mu = -\rho \left(\kappa^{\mathrm{a}} + \kappa^{\mathrm{s}}\right) R^{\mu \nu }u_{\nu} - \rho\left(\kappa^{\mathrm{s}} R^{\alpha \beta} u_{\alpha} u_{\beta} + \kappa^{\mathrm{a}} 4 \pi B\right) u^{\mu}~,
\label{eq:Galpha}
\end{equation}
where $\kappa = \kappa^\mathrm{a} + \kappa^\mathrm{s}$ is the grey (frequency-independent) opacity, with $\kappa^\mathrm{a}$ and $\kappa^\mathrm{s}$ being the contributions due to absorption and scattering, respectively, and $4\pi B = a_R T_\mathrm{gas}^4$ is the integrated blackbody (Planck) function at temperature $T_\mathrm{gas}$, with radiation constant $a_R=4\sigma/c$.

The full set of conservation equations to be solved can now be written as
\begin{eqnarray}
 \partial_t D + \partial_i (DV^i) &=& 0 ~,  \label{eqn:de} \\
\partial_t {\cal E} + \partial_i \left(-\sqrt{-g}~T^i_t\right) &=&
      -\sqrt{-g}~T^\alpha_\beta~\Gamma^\beta_{t \alpha} - \sqrt{-g}~G_t ~,
    \label{eqn:en2} \\
 \partial_t {\cal S}_j + \partial_i \left(\sqrt{-g}~T^i_j\right) &=&
      \sqrt{-g}~T^\alpha_\beta~\Gamma^\beta_{j \alpha} + \sqrt{-g}~G_j ~,
    \label{eqn:mom2} \\
 \partial_t {\cal R} + \partial_i \left(\sqrt{-g}~R^i_t\right) &=&
      \sqrt{-g}~R^\alpha_\beta~\Gamma^\beta_{t \alpha} - \sqrt{-g}~G_t ~,
    \label{eqn:rad_en} \\
 \partial_t {\cal R}_j + \partial_i \left(\sqrt{-g}~R^i_j\right) &=&
      \sqrt{-g}~R^\alpha_\beta~\Gamma^\beta_{j \alpha} - \sqrt{-g}~G_j ~,
    \label{eqn:rad_mom} 
\end{eqnarray}
where $D=W\rho$ is the generalized fluid density, $W=\sqrt{-g} u^t = \sqrt{-g}\gamma/\alpha$ is the relativistic boost factor, $V^i=u^i/u^t$ is the fluid transport velocity, ${\cal E} = -\sqrt{-g} T^t_t$ is the total energy density, ${\cal S}_j = \sqrt{-g} T^t_j$ is the covariant momentum density, and ${\cal R}$ and ${\cal R}_j$ are the conserved radiation fields already defined.  To proceed, we utilize a new hybrid explicit-implicit scheme, primarily intended to address stability issues associated with the radiation source term (Section \ref{sec:source}).  In the first step of this method, we use the explicit HRSC method, described in \citet{fragile12}, to update the set of conserved variables 
\begin{equation}
\mathbf{U} = \left( \begin{array}{c} D \\ {\cal E} \\ {\cal S}_j \\ {\cal R} \\ {\cal R}_j \end{array} \right)
\end{equation}
to an intermediate state based on the following finite volume representation
\begin{equation}
\mathbf{U}^* = \mathbf{U}^n - \frac{\Delta t}{V} \sum\limits_{faces}\left(\mathbf{F}^i A_i\right)^n + \Delta t ~\mathbf{S}_c^n ~,
\end{equation}
accounting for the curvature source terms
\begin{equation}
\mathbf{S}_c(\mathbf{P}) = \left( \begin{array}{c} 0 \\ -\sqrt{-g}~T^\alpha_\beta~\Gamma^\beta_{t \alpha} \\ \sqrt{-g}~T^\alpha_\beta~\Gamma^\beta_{j \alpha} \\ \sqrt{-g}~R^\alpha_\beta~\Gamma^\beta_{t \alpha} \\ \sqrt{-g}~R^\alpha_\beta~\Gamma^\beta_{j \alpha}  \end{array} \right) ~,
\end{equation}
and flux terms 
\begin{equation}
\mathbf{F}^i(\mathbf{P}) = \left( \begin{array}{c} DV^i \\ -\sqrt{-g}~T^i_t \\ \sqrt{-g}~T^i_j \\ \sqrt{-g}~R^i_t \\ \sqrt{-g}~R^i_j  \end{array} \right) ~.
\end{equation}
The flux terms are calculated at zone faces using either the Harten-Lax-van Leer (HLL) or Lax-Friedrichs Riemann solver with either linear or PPM slope limited reconstruction of the primitive fields.

\subsection{Radiation Source Term}
\label{sec:source}

Once the explicit step is complete, we follow it with an implicit one of the form 
\begin{equation}
\mathbf{U}^{n+1} = \mathbf{U}^* + \Delta t ~\mathbf{S}_r^{n+1} ~
\end{equation}
that attempts to complete the update by accounting for the radiation source terms 
\begin{equation}
\mathbf{S}_r(\mathbf{P}) = \left( \begin{array}{c} 0 \\ -\sqrt{-g}~G_t \\ \sqrt{-g}~G_j \\ -\sqrt{-g}~G_t \\ -\sqrt{-g}~G_j \end{array} \right) ~.
\end{equation}
We perform the implicit integration iteratively, with the $m+1$ guess given by
\begin{equation}
\mathbf{U}^{m+1} = \mathbf{U}^* + \Delta t ~\mathbf{S}_r^{m+1} ~.
\label{eqn:iterate}
\end{equation}
Taking the 1st order Taylor expansion of the first term in equation (\ref{eqn:iterate}) and the radiation 4-force density $G_\alpha$, with respect to the primitive variables, we can approximate the ($m+1$)st iterate as
\begin{eqnarray}
\mathbf{U}^{m+1} &=& \mathbf{U}^m + \sum_a \left(\frac{\partial\mathbf{U}}{\partial P^a}\right)^m \delta P^a \\
G_\alpha^{m+1} &=& G_\alpha^m + \sum_a \left(\frac{\partial G_\alpha}{\partial P^a}\right)^m \delta P^a ~,
\end{eqnarray}
where
\begin{eqnarray} 
\delta \mathbf{P} = \left(\begin{array}{c}
   \delta \rho \\
   \delta  \epsilon \\
   \delta \widetilde{u}^i \\
   \delta E_R \\
   \delta \widetilde{u}_R^i \end{array}\right)
=\left(\begin{array}{c}
   \rho^{m + 1} - \rho^m \\
   \epsilon^{m + 1} -  \epsilon^m \\
   (\widetilde{u}^i)^{m+1} - (\widetilde{u}^i)^{m} \\
   E_R^{m + 1} - E_R^m \\
   (\widetilde{u}_R^i)^{m+1} - (\widetilde{u}_R^i)^m \end{array}\right)
~.
\end{eqnarray}
Plugging the expanded form of each variable into equation (\ref{eqn:iterate}), we get the following set of equations for the primitive fields $\delta P^a$
\begin{equation}
\sum_a \left(\frac{\partial\mathbf{U}^m}{\partial P^a} - 
             \Delta t~\frac{\partial\mathbf{S_r}^m}{\partial P^a}\right)
       \delta P^a
       = \mathbf{U}^* - \left(\mathbf{U}^m - \Delta t ~\mathbf{S_r}^m\right) ~.
\end{equation}
We now have a set of linear equations that can be represented as a single matrix equation of the form
\begin{equation}
{\bf A}{\bf x} = {\bf b} ~,
\label{eqn:matrix}
\end{equation}
with Jacobian matrix
\begin{equation}
A_{ba} = \left(\frac{\partial U^b}{\partial P^a} - \Delta t ~\frac{\partial S_r^b}{\partial P^a}\right) ~,
\label{eqn:jacmatrix}
\end{equation}
or more explicitly
\begin{equation}
{\bf A} = \left( \begin{array}{ccccc} 
  \frac{\partial D}{\partial \rho} 
       & 0 
       & \frac{\partial D}{\partial \widetilde{u}^i} 
       & 0 
       & 0 \\ 
  \frac{\partial {\cal E}}{\partial \rho}                     + \Delta t \sqrt{-g}\frac{\partial G_t}{\partial \rho} 
       & \frac{\partial {\cal E}}{\partial \epsilon}          + \Delta t \sqrt{-g}\frac{\partial G_t}{\partial \epsilon} 
       & \frac{\partial {\cal E}}{\partial \widetilde{u}^i}   + \Delta t \sqrt{-g}\frac{\partial G_t}{\partial \widetilde{u}^i} 
       &                                                        \Delta t \sqrt{-g}\frac{\partial G_t}{\partial E_R} 
       &                                                        \Delta t \sqrt{-g}\frac{\partial G_t}{\partial \widetilde{u}_R^i} \\
  \frac{\partial {\cal S}_j}{\partial \rho}                   - \Delta t \sqrt{-g}\frac{\partial G_j}{\partial \rho} 
       & \frac{\partial {\cal S}_j}{\partial \epsilon}        - \Delta t \sqrt{-g}\frac{\partial G_j}{\partial \epsilon} 
       & \frac{\partial {\cal S}_j}{\partial \widetilde{u}^i} - \Delta t \sqrt{-g}\frac{\partial G_j}{\partial \widetilde{u}^i} 
       &                                                      - \Delta t \sqrt{-g}\frac{\partial G_j}{\partial E_R} 
       &                                                      - \Delta t \sqrt{-g}\frac{\partial G_j}{\partial \widetilde{u}_R^i} \\
                                                                \Delta t \sqrt{-g}\frac{\partial G_t}{\partial \rho} 
       &                                                        \Delta t \sqrt{-g}\frac{\partial G_t}{\partial \epsilon} 
       &                                                        \Delta t \sqrt{-g}\frac{\partial G_t}{\partial \widetilde{u}^i} 
       & \frac{\partial {\cal R}}{\partial E_R}               + \Delta t \sqrt{-g}\frac{\partial G_t}{\partial E_R} 
       & \frac{\partial {\cal R}}{\partial \widetilde{u}_R^i} + \Delta t \sqrt{-g}\frac{\partial G_t}{\partial \widetilde{u}_R^i} \\
                                                                \Delta t \sqrt{-g}\frac{\partial G_j}{\partial \rho} 
       &                                                        \Delta t \sqrt{-g}\frac{\partial G_j}{\partial \epsilon} 
       &                                                        \Delta t \sqrt{-g}\frac{\partial G_j}{\partial \widetilde{u}^i} 
       & \frac{\partial {\cal R}_j}{\partial E_R}             + \Delta t \sqrt{-g}\frac{\partial G_j}{\partial E_R} 
       & \frac{\partial {\cal R}_j}{\partial\widetilde{u}_R^i}+ \Delta t \sqrt{-g}\frac{\partial G_j}{\partial \widetilde{u}_R^i}
\end{array} \right) ~,
\label{eqn:A}
\end{equation}
with
\begin{equation}
{\bf x} = \delta\mathbf{P} = 
          \left( \begin{array}{c} \delta \rho \\ 
                                  \delta \epsilon \\ 
                                  \delta \widetilde{u}^i \\ 
                                  \delta E_R \\ 
                                  \delta \widetilde{u}_R^i \end{array} \right) ~,
\end{equation}
and
\begin{equation}
{\bf b} = \mathbf{U}^* - (\mathbf{U}^m - \Delta t ~\mathbf{S_r}^m)
        = \left( \begin{array}{c} D^* - D^m \\ {\cal E}^* - {\cal E}^m - \Delta t \sqrt{-g} G_t^m \\ {\cal S}_j^* - {\cal S}_j^m + \Delta t \sqrt{-g} G_j^m \\ {\cal R}^* - {\cal R}^m - \Delta t \sqrt{-g} G_t^m \\ {\cal R}_j^* - {\cal R}_j^m - \Delta t \sqrt{-g} G_j^m \end{array} \right) ~.
\end{equation}
Note that ${\bf A}$ is really a $9 \times 9$ matrix, and ${\bf x}$ and ${\bf b}$ are 9-dimensional vectors; we have simply condensed the notation by representing each 3-vector in ${\bf A}$, ${\bf x}$, and ${\bf b}$ as a single entry.  

The important point is that the matrix ${\bf A}$ and vector ${\bf b}$ only include terms known at iteration $m$.  From these we can solve for the vector of unknown primitives at iteration $m+1$, ${\bf P}^{m+1} = {\bf P}^m + {\bf x}$, by inverting the matrix ${\bf A}$ and solving for ${\bf x}$ in equation (\ref{eqn:matrix}).  For the initial $m=0$ guess, we use the values of $\mathbf{P}$ from the previous timestep.  At each step, the conserved variables $\mathbf{U}^m$ are recalculated from the corresponding primitive set $\mathbf{P}^m$.  We iterate until
\begin{equation}
\frac{{\bf x}}{\mathbf{P}^{m+1}} \le \mathrm{tol} ~
\end{equation}
or the number of iterations exceeds some maximum.  Typical values are $\mathrm{tol} = 10^{-6}$, with the maximum number of iterations being 20.  Only rarely is the maximum iteration count exceeded in the problems presented in this work.  In these rare cases, the code will try both the analytic and numerical matrix inversion procedures.  If both fail, then the code replaces the primitive quantities in the problematic cell with an average of surrounding neighbor cells that have not failed this step.  We have found in our testing that using a tighter tolerance imposes a small penalty in terms of computational time.  For example, tightening the tolerance to $10^{-10}$ adds about 6\% to the computational time for the radiation shock tube tests presented in Section \ref{sec:radtube}.  

We have tested both analytic and numerical procedures for calculating the derivatives in (\ref{eqn:A}).  Both give reasonable and consistent results; we presently use the analytic method as our first option, with the numerical one acting as a back-up.  Appendix \ref{sec:derivs} reports all of the necessary derivatives for calculating ${\bf A}$ analytically. The numerical method is based on a forward difference approximation for the Jacobian matrix (\ref{eqn:jacmatrix}), provided all conserved fields and source terms are evaluated as functions of the primitive iterates.

Since the matrix equation (\ref{eqn:matrix}) involves only local calculations, its evaluation does not suffer from the scaling difficulties known to plague global matrix inversions.  {\it Cosmos++} has a number of matrix solvers and preconditioners built in.
The ones that we found worked well in this application are LU decomposition and Gaussian elimination.  We tried each of these on the cloud shadow test problem in Section \ref{sec:cloud}, where we report what few differences we found.

This method of solving the radiation source terms clearly also accomplishes the primitive inversion step, since we ultimately end up with the set of primitives $\mathbf{P}$ at the new timestep $n+1$.  In fact, this approach is quite similar to the $5D$ primitive inversion scheme described in \citet{noble06}, which we have added as an option in {\it Cosmos++}, independent of whether or not the radiation package is being used.  The method is also similar to the failure recovery option described in Appendix A of \citet{sadowski13}.  We mention again that magnetic fields can easily be included since their primitive form is trivially related to their conserved form \citep[c.f.][]{fragile12} and they do not enter into the radiation source terms.

\subsection{Wave Speeds}
\label{sec:speeds}

One of the advantages of reformulating our GR radiation MHD method in terms of the new primitive radiation variables, $E_R$ and $u^i_R$, is that it makes it trivially easy to calculate the characteristic wave speeds associated with the radiation fields, as required for our approximate Riemann solvers.  In the radiation rest frame, this speed is simply $c/\sqrt{3}$.  Following a procedure similar to that described in Section 2.5 of \citet{fragile12}, we simply boost this speed into the grid (a.k.a. coordinate) frame.

\section{Test Problems}
\label{sec:tests}

In this section we present a series of test problems meant to validate various aspects of our new method.  Where appropriate, we include comparisons to our previous method.  

\subsection{Radiation Shock Tubes}
\label{sec:radtube}

The first tests we perform are the same four radiative shock tube tests first introduced in \citet{farris08} and repeated in \citet{zanotti11,fragile12,sadowski13}.  Each test includes a different nonlinear radiation-hydrodynamic wave, specifically:  a nonrelativistic strong shock (case 1); a mildly relativistic strong shock (case 2); a highly relativistic wave (case 3); and a radiation-pressure-dominated, mildly relativistic wave (case 4).  We initialize each test with a ``left'' and ``right'' state, initially separated by an imaginary partition.  The goal of these tests is two-fold: 1) to confirm that the new semi-implicit scheme can reproduce the results of our previous fully explicit scheme; and 2) to facilitate a straightforward comparison of the performance (primarily speed) of the two methods.  With regards to the first goal, the results of all four tests are visually indistinguishable from Figures 1-4 of \citet{fragile12}; therefore, we do not reproduce those figures here.  We also monitored the $L$-1 norm error (i.e. $\vert E(a) \vert_1 = \sum_i \Delta x \vert a_i - A_i \vert$, where $a_i$ and $A_i$ are the numerical and semi-analytic solutions, respectively) for the case 4 test.  The errors with the semi-implicit scheme were nearly identical to those reported in Table 2 of \citet{fragile12}, demonstrating that the convergence rate on this test is again almost exactly 2 when using piecewise linear interpolation, as it should be.

As for the second goal of testing the performance, we found, not surprisingly, that the semi-implicit scheme is somewhat slower than the fully explicit one.  On the case 3 test with 800 zones on a single 2.4 GHz processor on the College of Charleston cluster, the fully explicit method executed 43 cycles/s, while the semi-implicit scheme with the LU decomposition matrix solver only managed 23 cycles/s (both for $\text{tol} = 10^{-6}$).  However, because of the increased stability of the semi-implicit scheme, the new code is able to run this test with a timestep that is 10 times larger than that used by our previously published code, so we are still better off in terms of total CPU time.

\subsection{Radiation-Modified MHD Linear Waves}
The previous section demonstrates the convergence rate of the new numerical scheme in the presence of only optically thick radiation fields; here we extend the convergence tests to investigate the performance of the $\bf{M}_1$ closure over a broader range of optical depths and in cases that include stiff source terms. To do so, we reproduce a set of tests, initially performed in the Newtonian limit by \citet{jiang12} and then in relativistic form by \citet{sadowski14, mckinney14}, that require accurately propagating linear MHD waves in the presence of radiation. 

The tests include sound, as well as fast and slow magnetosonic, waves, all propagating within various optical depth backgrounds. Each test begins by initializing eigenmodes of the form 
\begin{equation}
q^a = Re\left[q^a_0 + \delta q^a e^{i\left( \omega t - k x \right)} \right] 
\label{eq:perturb}
\end{equation}
for each fluid and radiation variable, where $\delta q^a$ are eigenvectors given in Table \ref{tab:linearwave}, which is reproduced from \citet{mckinney14}. The $Re[..]$ represents the real part of a given variable. The unperturbed background gas and radiation fields have the following parameters: $\rho_0 = 1$, $u^x_0 = u^y_0 = 0$, and $F^x_0 = F^y_0 = 0$.  The sound speed in the background is $c_{s,0} = 0.1$. For a $\Gamma = 5/3$ gas, which we assume, this gives a gas internal energy of $e_0 = \rho_0\epsilon_0 = \rho_0[\Gamma(\Gamma-1) c_{s,0}^{-2} - \Gamma]^{-1} = 9.13706 \times 10^{-3}$.  In cases where magnetic fields are included, the Alfv\'{e}n speed is $v_{A,0} = 0.2$.  We split the background field evenly between $x$ and $y$ components, such that $B^x_0 = B^y_0 = 0.100759$.  The background radiation pressure is set from the dimensionless parameter, $\mathbb{P}$ = $P_\mathrm{rad,0}/P_\mathrm{gas,0}$, where the thermal pressure of the background gas is $P_\mathrm{gas,0} = (\Gamma - 1)e_0$ and the radiation energy in the fluid rest frame is $E_0 = 3 P_\mathrm{rad,0}$.  We wish to also have the radiation field start in local thermodynamic equilibrium with the gas, i.e. $E_0 = a_R T_\mathrm{gas,0}^4$, where $T_\mathrm{gas,0} = P_\mathrm{gas,0}/\rho_0$ (in units where $m_H/k_B = 1$).  This requires that we redefine the radiation constant, $a_R$, or equivalently the Stefan-Boltzmann constant, $\sigma$, for this problem.

\begin{deluxetable}{cccc}
\tabletypesize{\scriptsize}
\tablecaption{ Eigenmodes of Linear RMHD Waves \label{tab:linearwave}}
\tablewidth{0pt}
\tablehead{
\hline
\colhead{ } & \colhead{sound, $\tau = 0.1$, $\mathbb{P} = 0.1$} & \colhead{fast, $\tau = 0.1$, $\mathbb{P} = 0.1$} & \colhead{slow, $\tau = 0.1$, $\mathbb{P} = 0.1$}} 
\startdata
$\delta \rho$ & $10^{-6} + 0i$  & $10^{-6} + 0i$  & $10^{-6} + 0i$   \\
$\delta e$ & $1.51557 \times 10^{-8} + 7.69693 \times 10^{-10}i$ & $1.51984 \times 10^{-8} + 4.81575 \times 10^{-10}i$  & $1.50174 \times 10^{-8} + 1.22299 \times 10^{-9}i$  \\
$\delta \mathrm{u}^x$ & $9.97992 \times 10^{-8} + 2.55207 \times 10^{-9}i$ &  $1.60251 \times 10^{-7} + 7.23831 \times 10^{-10}i$  &  $6.15333 \times 10^{-8} + 1.83140 \times 10^{-9}i$   \\
$\delta \mathrm{u}^y$ & $0 + 0i$ &  $-9.79544 \times 10^{-8} + 9.83679 \times 10^{-10}i$   &  $9.89772 \times 10^{-8} + 6.54186 \times 10^{-9}i$   \\
$\delta \mathrm{B}^y$ & $0 + 0i$ & $1.62344 \times 10^{-7} - 8.96662 \times 10^{-10}i$  & $-6.14882 \times 10^{-8}  - 5.88315 \times 10^{-9}i$ \\
$\delta$E &$1.33148 \times 10^{-13} + 3.60017 \times 10^{-11}i$ & $1.48421 \times 10^{-12} + 6.06322 \times 10^{-11}i$ & $1.91703 \times 10^{-13} + 2.18721 \times 10^{-11}i$  \\
$\delta$F$^x$ & $-2.52471 \times 10^{-10} + 7.40041 \times 10^{-11}i$ & $-3.95433 \times 10^{-10} + 8.51051 \times 10^{-11}i$ &  $-1.65181 \times 10^{-10} + 7.17520 \times 10^{-11}i$ \\
$\delta$F$^y$ & $0 + 0i$ & $2.36680 \times 10^{-10} + 2.11182 \times 10^{-11}i$ &  $-2.23679 \times 10^{-10} - 7.43141 \times 10^{-11}i$ \\
$\delta \omega$ & $0.627057 + 0.0160351i$ & $1.00689 + 0.00454797i$ &  $0.386625 + 0.0115070i$ \\
\hline
\colhead{ } & \colhead{sound, $\tau = 10$, $\mathbb{P} = 0.1$} & \colhead{fast, $\tau = 10$, $\mathbb{P} = 0.1$} & \colhead{slow, $\tau = 10$, $\mathbb{P} = 0.1$} \\
\hline
$\delta \rho$ & $10^{-6} + 0i$  & $10^{-6} + 0i$  & $10^{-6} + 0i$   \\
$\delta e$ & $1.17977 \times 10^{-8} + 3.04292 \times 10^{-9}i$ & $1.31055 \times 10^{-8} + 2.26908 \times 10^{-9}i$ & $1.03536 \times 10^{-8} + 2.54899 \times 10^{-9}i$  \\
$\delta \mathrm{u}^x$ &  $9.29099 \times 10^{-8} + 1.44382 \times 10^{-8}i$ &  $1.59215 \times 10^{-7} + 4.26731 \times 10^{-9}i$  &  $5.51071 \times 10^{-8} + 6.81772 \times 10^{-9}i$   \\
$\delta \mathrm{u}^y$ & $0 + 0i$ &  $-9.86566 \times 10^{-8} + 6.11642 \times 10^{-9}i$   &  $7.68348 \times 10^{-8} + 1.80696 \times 10^{-8}i$   \\
$\delta \mathrm{B}^y$ & $0 + 0i$ & $1.63045 \times 10^{-7} - 5.54016 \times 10^{-9}i$  & $-4.16352 \times 10^{-8}  - 1.54221 \times 10^{-8}i$ \\
$\delta$E & $1.98198 \times 10^{-9} + 2.20605 \times 10^{-9}i$ & $2.95346 \times 10^{-9} + 1.59681 \times 10^{-9}i$ & $9.05892 \times 10^{-10} + 1.85949 \times 10^{-9}i$  \\
$\delta$F$^x$ & $-4.36777 \times 10^{-10} + 4.31621 \times 10^{-10}i$ & $-2.72196 \times 10^{-10} + 6.08654 \times 10^{-10}i$ &  $-3.83041 \times 10^{-10} + 1.99268 \times 10^{-10}i$ \\
$\delta$F$^y$ & $0 + 0i$ & $3.23863 \times 10^{-12} + 2.38271 \times 10^{-11}i$ &  $2.11406 \times 10^{-12} - 6.39415 \times 10^{-12}i$ \\
$\delta \omega$ & $0.583770 + 0.0907181i$ & $1.00038 + 0.0268123i$ &  $0.346248 + 0.0428370i$ \\
\hline
\colhead{ } & \colhead{sound, $\tau = 0.1$, $\mathbb{P} = 10$} & \colhead{fast, $\tau = 0.1$, $\mathbb{P} = 10$} & \colhead{slow, $\tau = 0.1$, $\mathbb{P} = 10$} \\
\hline
$\delta \rho$ & $10^{-6} + 0i$  & $10^{-6} + 0i$  & $10^{-6} + 0i$   \\
$\delta e$ & $9.14134 \times 10^{-9} + 3.83221 \times 10^{-10}i$ & $9.20593 \times 10^{-9} + 7.43789 \times 10^{-10}i$ & $9.13450 \times 10^{-9} + 2.48194 \times 10^{-10}i$  \\
$\delta \mathrm{u}^x$ &  $7.74805 \times 10^{-8} + 3.58319 \times 10^{-9}i$ &  $1.51648 \times 10^{-7} + 2.92431 \times 10^{-9}i$  &  $5.01905 \times 10^{-8} + 2.38662 \times 10^{-9}i$   \\
$\delta \mathrm{u}^y$ & $0 + 0i$ &  $-1.11472 \times 10^{-7} + 6.59618 \times 10^{-10}i$   &  $6.76529 \times 10^{-8} + 3.82639 \times 10^{-9}i$   \\
$\delta \mathrm{B}^y$ & $0 + 0i$ & $1.74787 \times 10^{-7} - 1.86580 \times 10^{-9}i$  & $-3.51141 \times 10^{-8}  - 1.22066 \times 10^{-9}i$ \\
$\delta$E & $-1.52193 \times 10^{-10} + 9.38041 \times 10^{-10}i$ & $-4.70213 \times 10^{-10} + 1.98234 \times 10^{-9}i$ & $-7.35475 \times 10^{-11} + 6.00745 \times 10^{-10}i$  \\
$\delta$F$^x$ & $-1.93666 \times 10^{-8} - 7.93047 \times 10^{-10}i$ & $-3.79422 \times 10^{-8} - 3.18097 \times 10^{-10}i$ &  $-1.25407 \times 10^{-8} - 5.53622 \times 10^{-10}i$ \\
$\delta$F$^y$ & $0 + 0i$ & $2.69349 \times 10^{-8} + 2.67047 \times 10^{-9}i$ &  $-1.48948 \times 10^{-8} - 5.73105 \times 10^{-9}i$ \\
$\delta \omega$ & $0.486824 + 0.0225139i$ & $0.952830 + 0.0183740i$ &  $0.315356 + 0.0149956i$ \\
\hline
\colhead{ } & \colhead{sound, $\tau = 10$, $\mathbb{P} = 10$} & \colhead{fast, $\tau = 10$, $\mathbb{P} = 10$} & \colhead{slow, $\tau = 10$, $\mathbb{P} = 10$} \\
\hline
$\delta \rho$ & $10^{-6} + 0i$  & $10^{-6} + 0i$  & $10^{-6} + 0i$   \\
$\delta e$ & $1.17070 \times 10^{-8} + 1.88153 \times 10^{-9}i$ & $1.17305 \times 10^{-8} + 1.71290 \times 10^{-9}i$  & $9.46189 \times 10^{-9} + 1.21376 \times 10^{-9}i$  \\
$\delta \mathrm{u}^x$ & $2.66251 \times 10^{-7} + 6.33514 \times 10^{-8}i$ &  $2.78499 \times 10^{-7} + 5.23804 \times 10^{-8}i$  &  $8.34269 \times 10^{-8} + 1.20829 \times 10^{-8}i$   \\
$\delta \mathrm{u}^y$ & $0 + 0i$ &  $-2.81093 \times 10^{-8} + 6.25588 \times 10^{-9}i$   & $1.13633 \times 10^{-7} + 2.72697 \times 10^{-7}i$   \\
$\delta \mathrm{B}^y$ & $0 + 0i$ & $1.10170 \times 10^{-7} - 4.03337 \times 10^{-9}i$  & $-8.03823 \times 10^{-8}  - 3.03114 \times 10^{-7}i$ \\
$\delta$E & $2.05419 \times 10^{-7} + 1.49859 \times 10^{-7}i$ & $2.07294 \times 10^{-7} + 1.36364 \times 10^{-7}i$ & $2.59666 \times 10^{-8} + 9.67891 \times 10^{-8}i$  \\
$\delta$F$^x$ & $-2.07308 \times 10^{-8} + 3.77556 \times 10^{-8}i$ & $-1.83331 \times 10^{-8} + 3.63664 \times 10^{-8}i$ &  $-1.98263 \times 10^{-8} + 5.47610 \times 10^{-9}i$ \\
$\delta$F$^y$ & $0 + 0i$ & $2.67581 \times 10^{-10} +1.24272 \times 10^{-9}i$ & $3.66075 \times 10^{-9} - 1.14750 \times 10^{-9}i$ \\
$\delta \omega$ & $1.67290 + 0.398049i$ & $1.74986 + 0.329116i$ &  $0.524187 + 0.0759190i$ \\
\enddata
\end{deluxetable}

This problem evolves on a one-dimensional grid covering the domain $0 \leq x \leq 1$, with periodic boundary conditions.  Note, though, that we evolve both the $x$ and $y$ components of vector quantities.  The wavenumber in equation (\ref{eq:perturb}) is taken to be $k=2\pi$, such that exactly one full wavelength fits on the computational domain. The gas interacts with the radiation through the absorption opacity, which is set equal to the optical depth of the domain, i.e. $\kappa^\mathrm{a} = \tau$, while the scattering opacity is set to zero\footnote{There is an error in the description of this problem in both \citet{sadowski14} and \citet{mckinney14}. Both state that $\kappa^\mathrm{s} = \tau$ and $\kappa^\mathrm{a} = 0$, while, in fact, it is the opposite.  Also, neither mentions that the radiation constant must be redefined.}, $\kappa^\mathrm{s} = 0$. 

\begin{figure}
\begin{center}
\includegraphics[width=0.9\columnwidth]{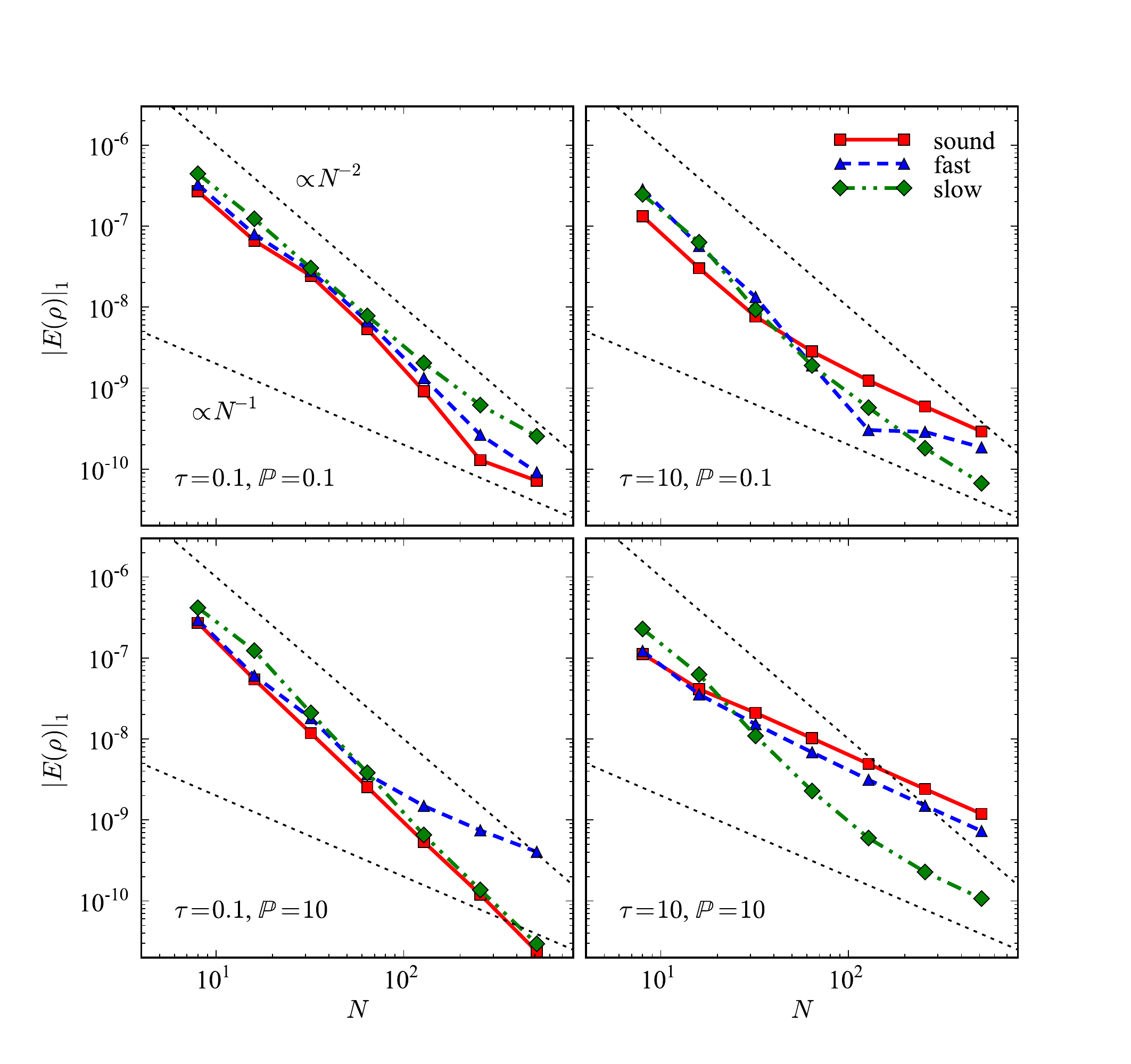} 
\caption{Plots of the $L$-1 norm error of density for the radiation-modified MHD linear wave tests.  $N$ is the number of cells used, ranging from 8 to 512.  The {\em dashed} lines show 1st and 2nd order convergence.  The optical depth and pressure ratio for each case are provided in the panels.
\label{fig:linearwave}}
\end{center}
\end{figure}

The problem is evolved for a time $t = 2\pi/Re[\omega]$, such that the wave should propagate back to its original location.  Because of the interaction of the gas with the radiation, the wave is damped by an amount $\exp(-Im[\omega] t)$, where $Im[..]$ refers to the imaginary part of the variable.  We test the convergence of our code by calculating the $L$-1 norm error of $\rho$, by comparing the numerical solution with the analytic one provided by (\ref{eq:perturb}).  Figure \ref{fig:linearwave} shows the results for resolutions from 8 to 512 zones.  In most cases, when optical depths are small, we get 2nd order convergence.  For large optical depths, on the other hand, while the convergence starts off at 2nd order at low resolutions, it switches to 1st order as the resolution increases.  This is where the new implicit treatment of the radiation source term dominates the error.

\subsection{Cloud Shadow}
\label{sec:cloud}

Among the many unphysical attributes of the Eddington approximation used by our previous code is that opaque objects cast no shadows.  This is because the radiation simply diffuses around the opaque object and fills in the ``shadow'' region.  The $\bf{M}_1$ closure used in our new method is expected to do much better in this regard.  To test this, we reproduce a classic problem originally introduced by \citet{hayes03}. The problem involves an opaque, spheroidal cloud embedded within a cylindrical box of transparent, low density gas, with a light source placed at one end of the cylinder.  The cylinder has length $L= 1$ cm along the $z$-axis, resolved with 280 zones, and radius $R=0.12$ cm, resolved with 80 zones.  The cloud lies in the center of the box at $(R_c, z_c) = (0, 0.5~\mathrm{cm})$ and its extension (Gaussian effective width) is $(R_0, z_0) = (0.06~\mathrm{cm}, 0.1~\mathrm{cm})$.  The gas and radiation begin in equilibrium with $T_\mathrm{gas} = T_\mathrm{rad} = 290~\mathrm{K}$, and the gas has an adiabatic index of $\Gamma = 5/3$. The density of the background gas is $\rho_0 = 1$ g cm$^{-3}$, while that of the cloud is $\rho_c = 1000$ g cm$^{-3}$.  The density of the cloud drops off exponentially at its surface as
\begin{equation}
\rho(z,r) = \rho_0 + \frac{\rho_c - \rho_0}{1 + \exp \Delta} ~,
\end{equation}
where 
\begin{equation}
\Delta = 10\left[\left(\frac{z - z_c}{z_0}\right)^{2} + \left(\frac{R - R_c}{R_0}\right)^{2} -1\right] ~.
\end{equation}
The opacity of the gas is assumed to come from thermal bremsstrahlung:
\begin{equation} 
 \kappa^{\mathrm{a}} = 0.1 \left(\frac{T}{T_{0}}\right)^{-7/2} \left(\frac{\rho}{\rho_0}\right) \mathrm{cm}^{2}~\mathrm{g}^{-1} ~,
\end{equation}
with no scattering contribution, $\kappa^{\mathrm{s}}=0$.  At the bottom boundary, a uniform source with $T_{\mathrm{source}}$ = 1740 K illuminates the cylinder.  The radiation fields are set to $E = a_R T^4$ and $F^z = 0.99999 E$.  Initially, the mean free path of the gas in the cylinder is 10 cm, whereas, within the cloud, it is $10^{-5}$ cm. This discrepancy in the mean free paths creates a shadow behind the cloud, which should remain stable until the light ultimately diffuses through the cloud.  

Before showing results using our new semi-implicit scheme and $\bf{M}_1$ closure, it is worth mentioning that this test could not even be performed with our original, fully-explicit radiation hydrodynamics scheme.  This is because instabilities in the optically-thin background gas would grow catastrophically within only a few cycles.  

\begin{figure}
\begin{center}
\includegraphics[width=0.33\columnwidth]{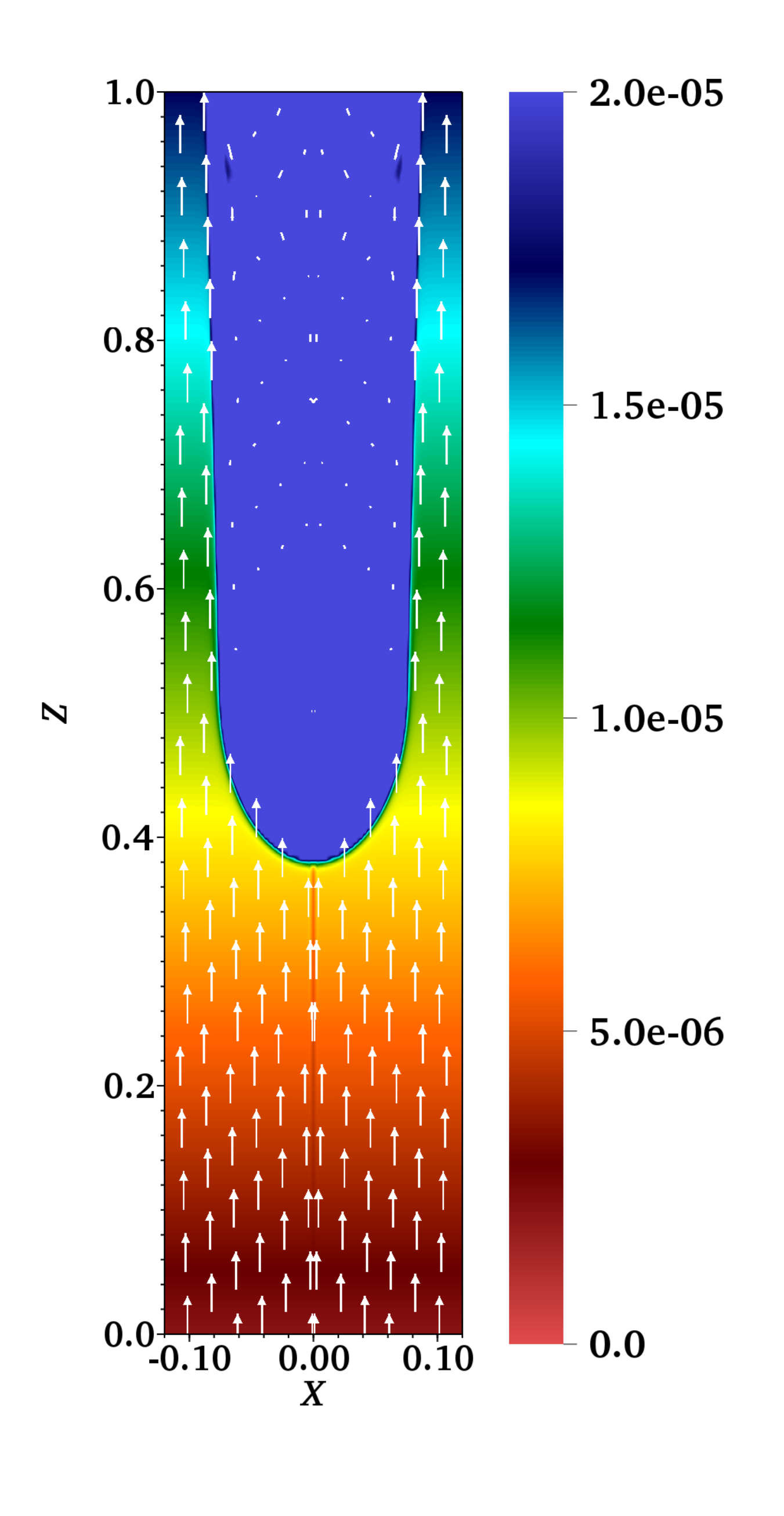} 
\caption{Pseudocolor plot of $E_R$ in units erg/cm$^3$ and vectors representing $u_R^i$ for the cloud shadow test after $10^{-10}$ s.
\label{fig:cloud}}
\end{center}
\end{figure}

The results with the new semi-implicit scheme and $\bf{M}_1$ closure are shown in Figure \ref{fig:cloud}.  As expected, the cloud produces a sharp, clear shadow; its edges gradually flare out, which is common with the $\bf{M}_1$ closure, but the transition from light to dark is nevertheless quite sharp.

Before moving on, we mention that we also used this test to compare the LU decomposition and Gaussian elimination linear solver options in {\em Cosmos++}.  We found they both produce qualitatively similar results at similar computational expense.  For example, on this cloud shadow test, run with $\mathrm{tol} = 10^{-6}$ on eight, 2.9 GHz processors on the College of Charleston cluster, the LU decomposition case ran to completion (5025 cycles) in 6590 s, while the Gaussian elimination case needed 6627 s.  Those times were with the analytic form of the 9D implicit solver; we also tested the numerical form and found it to be about 15\% slower.

Although this test demonstrates that the $\bf{M}_1$ closure is clearly an advance over the Eddington approximation used in our previous work, it is by no means without flaws of its own.  One well-known shortcoming is that beams of light can not cross one another in the $\bf{M}_1$ scheme.  Instead, intersecting beams merge, flowing in the direction of their resultant flux.  An example of this is shown in the two-beam test of \citet{jiang12, sadowski13}. Following \citet{sadowski13}, we set this problem up in units where $c = m_H/k_B = 1$, such that the ideal gas law is $P = \rho T$.  For this problem, we set up a Cartesian box spanning $-6 \le x \le 3$ and $0 \le y \le 1.5$, resolved with $100\times 50$ zones.  Constant boundary conditions are used at the left and top boundaries, from which a beam with $F^x = 0.93 E$ and $F^y = -0.37 E$ (with the magnitude $\vert F \vert$ restricted to be $0.999 E$) emanates.  A reflecting boundary is imposed at $y=0$, such that a virtual second beam intercepts the first along this boundary.  

Similar to the previous test, a cloud of radius 0.22 is placed at the origin of the coordinate system.  The cloud has a density $\rho_c = 1000$, while the background is $\rho_0 = 10^{-4}$, and the adiabatic index is $\Gamma = 1.4$.  At the boundary, the source temperature is $T_\mathrm{source} = 100$, while the background has $T_\mathrm{gas} = T_\mathrm{rad} = 1$. The radiation energy, as before, is set by $E = a_R T^4$. In this problem, we used a constant absorption opacity of magnitude $\kappa^{\mathrm{a}} = 1$.

\begin{figure}
\begin{center}
\includegraphics[width=0.9\columnwidth]{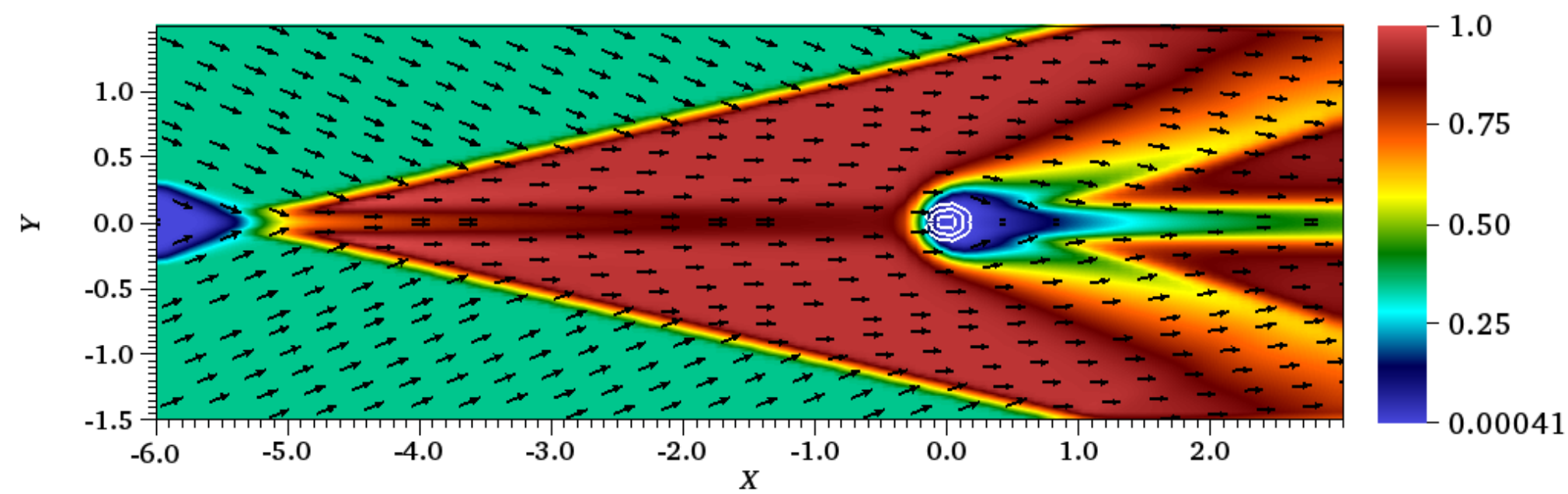} 
\caption{Pseudocolor plot of $-R^t_t$, with contours of $\rho$ at 50, 112, and 250, and vectors representing $u_R^i$ for the two-beam cloud shadow test at $t = 20$.  Note that we have reflected this image through the $y=0$ plane.
\label{fig:2beam}}
\end{center}
\end{figure}

Figure \ref{fig:2beam} shows the results of this two-beam test.  In the upper and lower left regions of this figure, we see the undisturbed beams entering the simulation domain.  To the right of these, we see the triangular region where the two beams overlap and interact with one another.  We see that the resulting flux is directed in the $x$-direction as expected.  However, although this now looks like a single, uniform beam approaching the cloud, it is quantitatively different than the beam in the previous case.  The main difference is that, while the previous case had $F^x \approx E$, in this case $F^x$ only equals $0.93 E$.  This means that the gradient of the specific intensity in the direction of the flux is not steep enough to ensure that all photons move in that direction.  Some photons still move in the direction of the original beams.  As a result, the cloud shadow no longer exhibits sharp edges as it did before.  We also see that the $\bf{M}_1$ closure produces a narrow shadow along the $x$-axis to the right of the cloud that cuts through what should be a uniform penumbral region.  Similar to \citet{mckinney14}, we find the edges of the beams to be sensitive to the reconstruction method and interpolation order used.  The results in Figure \ref{fig:2beam} use piecewise linear interpolation and the MINMOD slope limiter.  Tests with other options showed much more pronounced oscillations along the edges, oscillations which would penetrate into the beam in some cases.

Although this test highlighted a significant weakness of the $\bf{M}_1$ method, we do not expect it to greatly impact our main anticipated application of black hole accretion.  In accretion disks, multiple light sources are not expected to be encountered very often, except perhaps along the rotation axis.  In any case, the $\bf{M}_1$ closure is clearly an improvement over our previous method.

\subsection{Beam of Light Near a Black Hole}
\label{sec:beam}

As a test of radiation in a strong gravitational field, we reproduce one of the tests from \citet{sadowski13}, namely that of a beam of light near a black hole.  This particular test focuses on the propagation of a beam in the curved spacetime geometry near a $3 M_\odot$ Schwarzschild black hole.  Coupling between the gas and the radiation is neglected ($\kappa^\mathrm{a} = \kappa^\mathrm{s} = 0$, therefore $G^\mu = 0$).  Because {\em Cosmos++} has the capability to evolve the hydrodynamic and radiation fields separately, we evolve only the radiation fields in this test; all hydrodynamic variables are neglected.  Formally, this test takes place in the free-streaming limit, as the optical depth is zero.  However, as mentioned in Section \ref{sec:primitives}, our method for the radiation can not, strictly speaking, be used in this limit.  Instead, we must restrict the radiation rest frame velocity to be less than $c$.  For this test, we require the Lorentz factor of the radiation to be $\le 10$.

We perform the same three beam tests as \citet{sadowski13}.  The models are run on a two-dimensional, $r-\phi$ grid, with resolution $30 \times 60$ and grid coverage over $r_\mathrm{in} < r < r_\mathrm{out}$ and $0 \le \phi \le \pi/2$, with $r_\mathrm{in}$ and $r_\mathrm{out}$ given for each model in Table \ref{tab:beam}.  The beams are initially centered at the positions $r_\mathrm{beam}$, with widths given in Table \ref{tab:beam}.  Note that the beam in case 1 is centered at the photon orbit radius, $r_\mathrm{beam} = r_\mathrm{p.o.} = 3$, meaning that photons in the center of the beam should be able to orbit the black hole indefinitely.

\begin{deluxetable}{cccc}
\tabletypesize{\scriptsize}
\tablecaption{Light Beam Tests \label{tab:beam}}
\tablewidth{0pt}
\tablehead{
\colhead{Case} & \colhead{$r_\mathrm{beam}$} & \colhead{$r_\mathrm{in}$} & \colhead{$r_\mathrm{out}$} }
\startdata
1 & $3.0\pm0.1$ & 2.5 & 3.5  \\
2 & $6.0\pm0.2$ & 5.3 & 7.5  \\
3 & $16.0\pm0.5$ & 14.0 & 20.5
\enddata
\end{deluxetable}

The radiation temperature within the initial beam is $T_\mathrm{beam} = 10 T_0 = 10^7$ K, where $T_0$ is the temperature of the background radiation.  The radiation beam has an initial Lorentz factor of $\gamma = 10$ in the grid frame.  The beam initial conditions are held constant at the $\phi = 0$ boundary.

\begin{figure}
\begin{center}
\includegraphics[width=0.32\columnwidth]{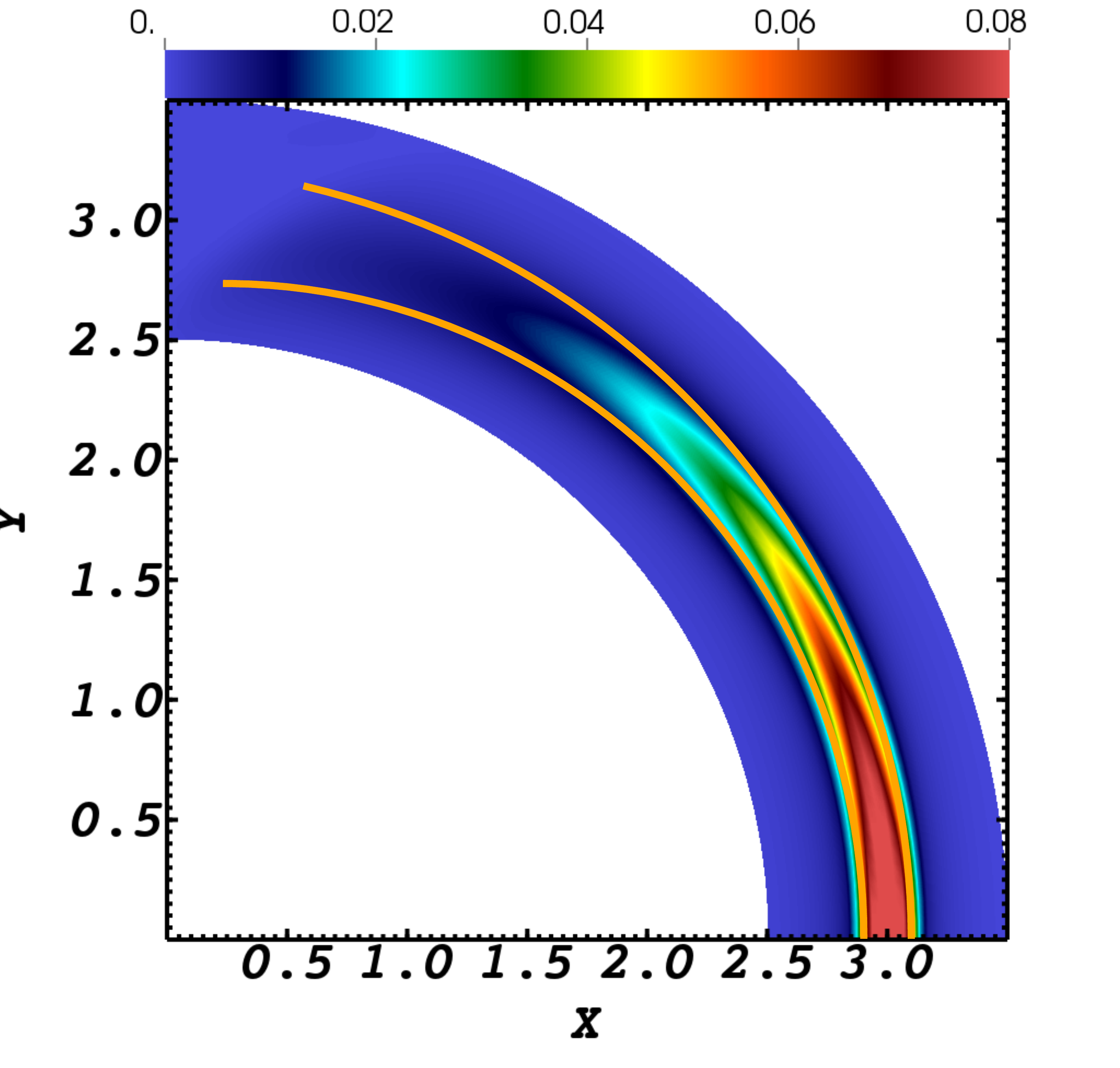} 
\includegraphics[width=0.32\columnwidth]{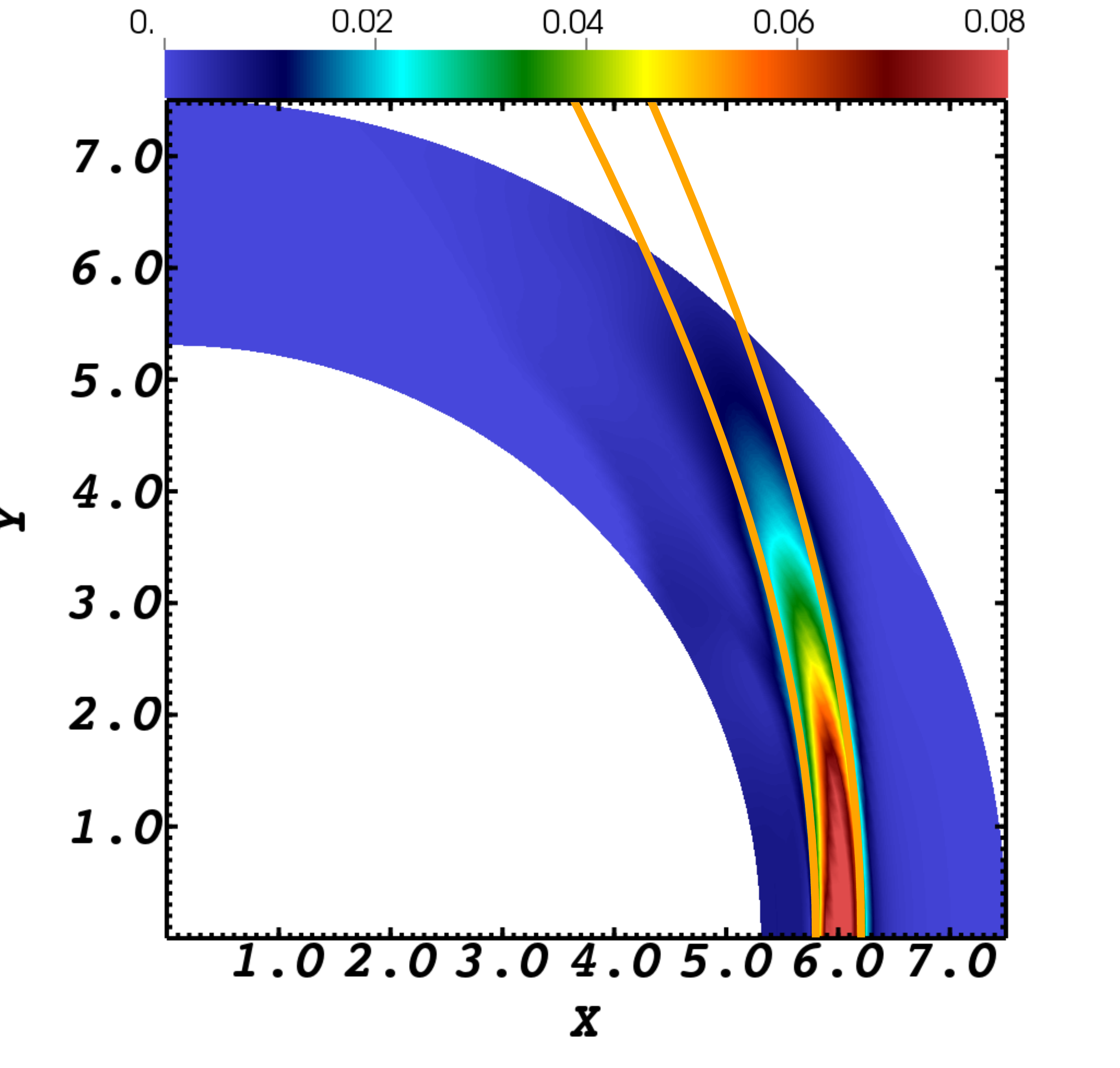} 
\includegraphics[width=0.32\columnwidth]{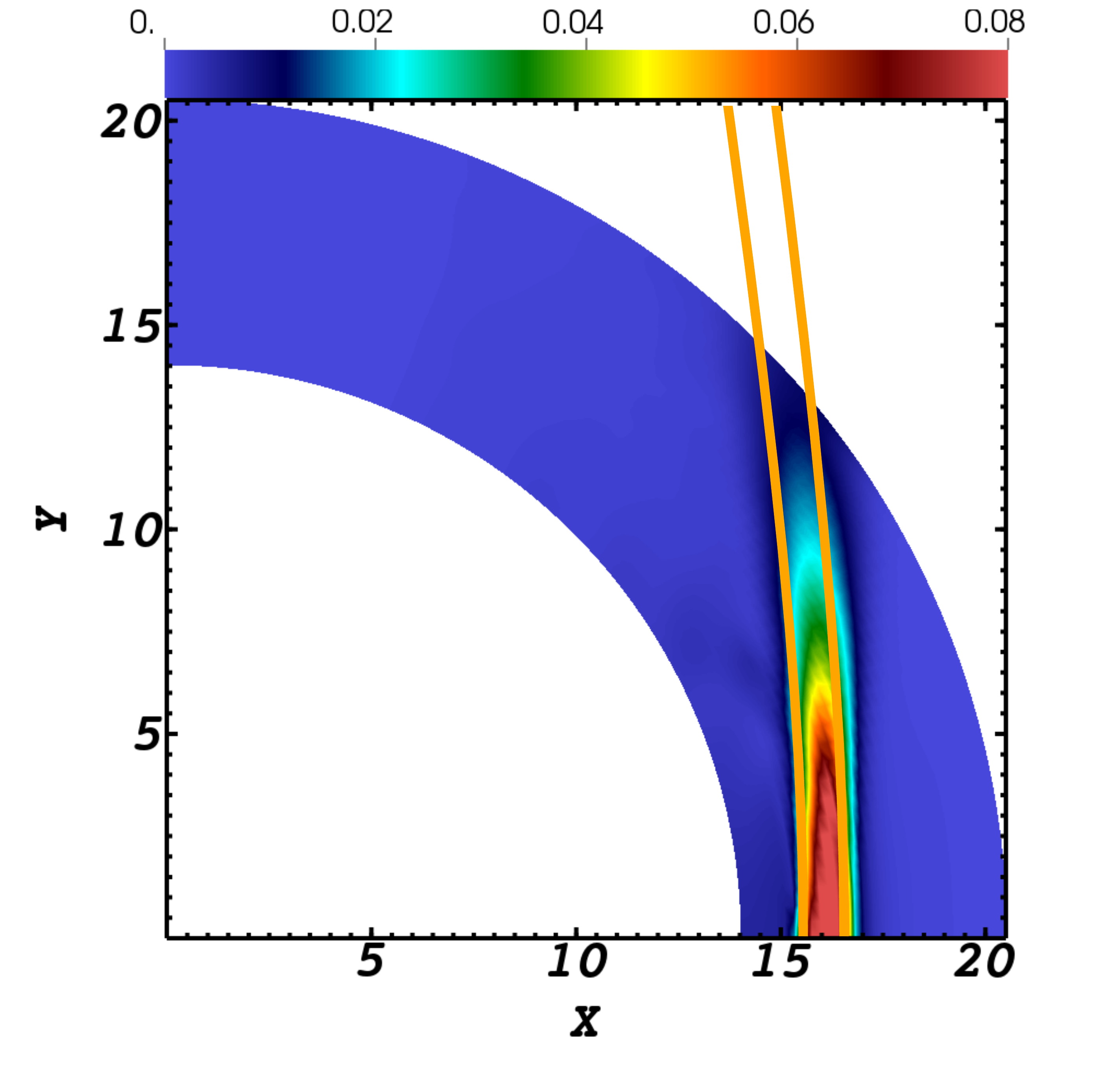} 
\caption{Pseudocolor of $E_R$ (in code units) for each case of the beam of light test.  A Schwarzschild black hole is located at the origin.  The orange curves represent geodesic paths starting at the initial inner and outer boundaries of the beam.
\label{fig:beam}}
\end{center}
\end{figure}

Figure \ref{fig:beam} shows the track of each radiation beam along with geodesic paths corresponding to the initial inner and outer boundaries of each beam.  We see that each beam experiences the expected curvature. 

\subsection{Static Atmosphere}

This test from \citet{sadowski13} considers a static atmosphere above a stellar surface.  This is done in the optically thin limit with $\kappa^\mathrm{a} = 0$ and $\kappa^\mathrm{s} = 0.4 ~\mathrm{cm}^2~\mathrm{g}^{-1}$.  The atmosphere is initially set up in hydrostatic equilibrium ($\partial_t = 0$ and $V^i_0 = 0$) such that 
\begin{equation}
\frac{1}{\rho_0} \frac{dP_0}{dr} = - \frac{1-f}{r^2} ~,
\label{eqn:dPdr}
\end{equation}
where $f = \kappa^\mathrm{s} F^r_\mathrm{in} r^2_\mathrm{in}$ and $F^r_\mathrm{in}$ is the radiative flux applied at the lower boundary of the atmosphere, $r_\mathrm{in}$.  As such, $f$ gives the ratio of the radiative to gravitational forces, with $f=1$ corresponding to the Eddington limit.

Assuming a polytropic equation of state, $P = K \rho^\Gamma$, equation (\ref{eqn:dPdr}) can be integrated to give 
\begin{equation}
\rho_0 = \left[ \frac{\Gamma-1}{\Gamma K} \left(C + \frac{1 - f}{r} \right) \right]^{1/(\Gamma-1)} ~,
\end{equation}
where
\begin{equation}
C = \frac{\Gamma K}{\Gamma - 1} \rho^{\Gamma-1}_\mathrm{in} - \frac{1-f}{r_\mathrm{in}} ~,
\end{equation}
where $\rho_\mathrm{in}$ is the rest mass density at $r_\mathrm{in}$.  In addition, energy conservation requires $F^r = F^r_\mathrm{in} r^2_\mathrm{in}/r^2$.  As was done in \citet{sadowski13}, we set $\rho_\mathrm{in} = 10^{-15}~\text{g~cm}^{-3}$ and $T_\mathrm{in} = 10^6~\text{K}$, which can be used to determine $P_\mathrm{in}$ and $K$.  The atmosphere extends from $r_\mathrm{in} = 10^6 r_G$ to $1.4 \times 10^6 r_G$, resolved with 40 grid zones, spaced linearly, where $r_G = GM/c^2$ is the gravitational radius; we fix this scale by setting $M = 1M_\odot$, such that $r_G = 15$ km.  The background geometry is set by the Schwarzschild metric.  The radiation energy is initially fixed to $E = F_\mathrm{in}/0.99$ (in the fluid frame).  We consider four input luminosities: $10^{-10}$, 0.1, 0.5, and $1.0 L_\mathrm{Edd}$.  Each case was run to a time of $t = 2 \times 10^9 M = 2.7$ hr.  

\begin{figure}
\begin{center}
\includegraphics[width=0.5\columnwidth]{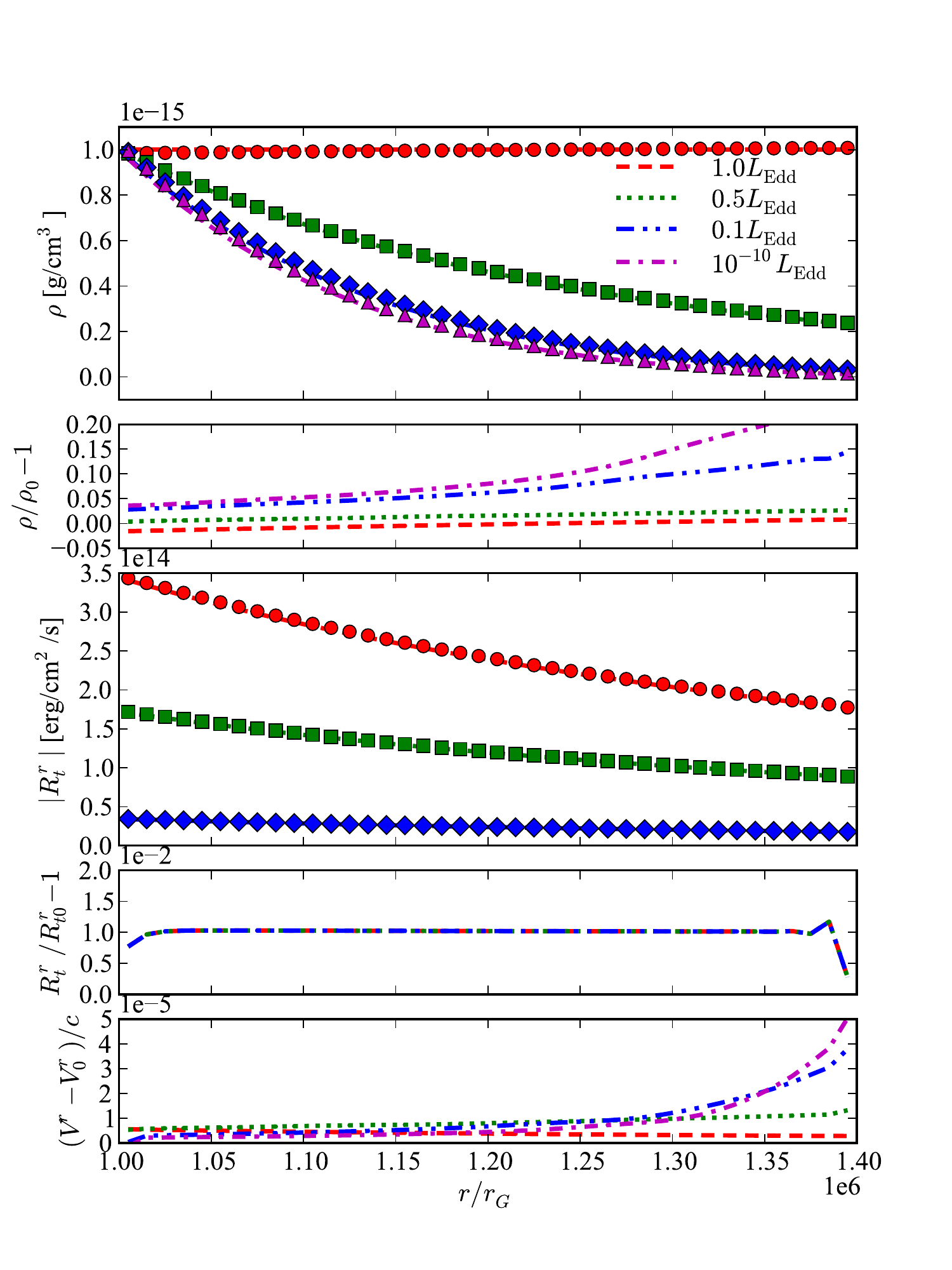}
\caption{Profiles of density, $\rho$, and radiation energy flux, $|R^r_t|$, along with errors $\rho/\rho_0-1$, $R^r_t/(R^r_t)_0 - 1$, and $(V^r-V^r_0)/c$, for the static atmosphere test.  Four different luminosities were considered.  Note the multiplicative scale factors for each variable, included near the positive end of each axis.  In the {\em first} and {\em third} panels, the symbols represent the final data, while the lines represent the initial conditions.
\label{fig:atmosphere}}
\end{center}
\end{figure}

Profiles of $\rho$ and $R^r_t = 4/3 E_R u^r_R (u_R)_t$, along with the errors $\rho/\rho_0-1$, $R^r_t/(R^r_t)_0 - 1$, and $(V^r-V^r_0)/c$, are shown in Figure \ref{fig:atmosphere}.  All of the numerical solutions lie reasonably close to the analytic ones, with errors mostly below a few percent.  We note, however, that our errors are considerably higher than those reported by \citet{sadowski13}.  This almost certainly has to do with the fact that they used a 5th order polynomial reconstruction scheme, whereas we used a linear (2nd order) one.

\subsection{Bondi Inflow with Radiation}
\label{sec:bondi}

In \citet{fragile12}, the case of optically-thick, spherical accretion onto a non-rotating black hole was considered using our fully explicit scheme. Here we revisit the problem using the hybrid implicit-explicit scheme and accounting for optically-thick and thin regions with the $\bf{M}_1$ closure. The setup of the problem follows \citet{fragile12}: We first fix the mass of the black hole, $M = 3 M_\odot$, and the density, $\rho_o$, and temperature, $T_o$, of the gas at the outer radius, $r_o = 10^4 r_\mathrm{S}$, where $r_\mathrm{S} = 2 r_G = 8.9$ km is the Schwarzschild radius.  Once $T_o$ and $\rho_o$ are fixed, and assuming some relation between $T_\mathrm{gas}$ and $T_\mathrm{rad}$, we can determine the polytropic index of the gas at $r_o$ from
\begin{equation}
\Gamma = 1 + \frac{1}{3} \left(\frac{P_\mathrm{gas} + P_\mathrm{rad}}{P_\mathrm{gas}/2 + P_\mathrm{rad}} \right) ~.
\end{equation}
We assume the initial value of $\Gamma$ found at $r_o$ applies throughout the flow for the duration of the simulations.  For the chosen parameters, this turns out to be $\Gamma = 5/3$.  By assuming a polytropic equation of state, $P \propto \rho^\Gamma$, we can determine the initial temperature profile of the gas from 
\begin{equation}
T_\mathrm{gas} = T_o (\rho/\rho_o)^{\Gamma-1} ~.
\label{eqn:T1}
\end{equation}
We still need to specify the initial profiles of $u^r$ and $\rho$.  For simplicity, we assume that these equal their free-fall values $u^r = -\sqrt{2M/r}$ and $\rho = -\dot{M}/4\pi r^2 u^r$ at all radii, with the mass accretion rate, $\dot{M}$, now one of our free parameters.  We explore mass accretion rates in the range $1 \le \dot{m} \le 100$, where $\dot{m} = \dot{M}/\dot{M}_\mathrm{Edd}$ and $\dot{M}_\mathrm{Edd}$ is the Eddington mass accretion rate.  This is another case where the new method in this paper has proven superior to our previous method, since $\dot{m} < 10$ was not achievable in \citet{fragile12}.  The grid for this problem uses a logarithmic radial coordinate of the form $x_1 \equiv 1+\ln(r/r_\mathrm{S})$, covering the spatial range $0.95 r_\mathrm{S} \le r \le r_o$.  All simulations use a one-dimensional grid with a resolution of 512 zones.  

During the evolution, the gas is allowed to interact with the radiation via two physical cooling processes: Thomson scattering and thermal bremsstrahlung.  The first contributes an opacity 
\begin{equation}
\kappa^\mathrm{s} = 0.4 ~\mathrm{cm}^2~\mathrm{g}^{-1} ~,
\end{equation}
while the second has the form  \citep{rybicki86} 
\begin{equation}
\kappa^\mathrm{a} = 1.7\times 10^{-25} T^{-7/2}_\mathrm{K} \rho_\mathrm{cgs} m^{-2}_\mathrm{p} ~\mathrm{cm}^2~\mathrm{g}^{-1} ~,
\end{equation}
where $T_\mathrm{K}$ is the ideal gas temperature of the fluid in Kelvin, $\rho_\mathrm{cgs}$ is the density in g/cm$^3$, and $m_\mathrm{p}$ is the mass of a proton in g.  We assume the gas is fully ionized hydrogen, so the mean molecular weight is $\mu = 0.5$.  In setting up the problem, we initially specify the ratio of radiation to gas temperature, $T_\mathrm{rad}/T_\mathrm{gas} \ll 1$.  This is mainly done so that the radiation energy density $E= a T^4_\mathrm{rad}$ may start with some reasonable value.  The radiation 4-velocity $u_R^r$ is initially set equal to the fluid 4-velocity $u^r$.  We confirm that our final results are not sensitive to our choices for these parameters.  

\begin{figure}
\begin{center}
\includegraphics[width=0.5\columnwidth]{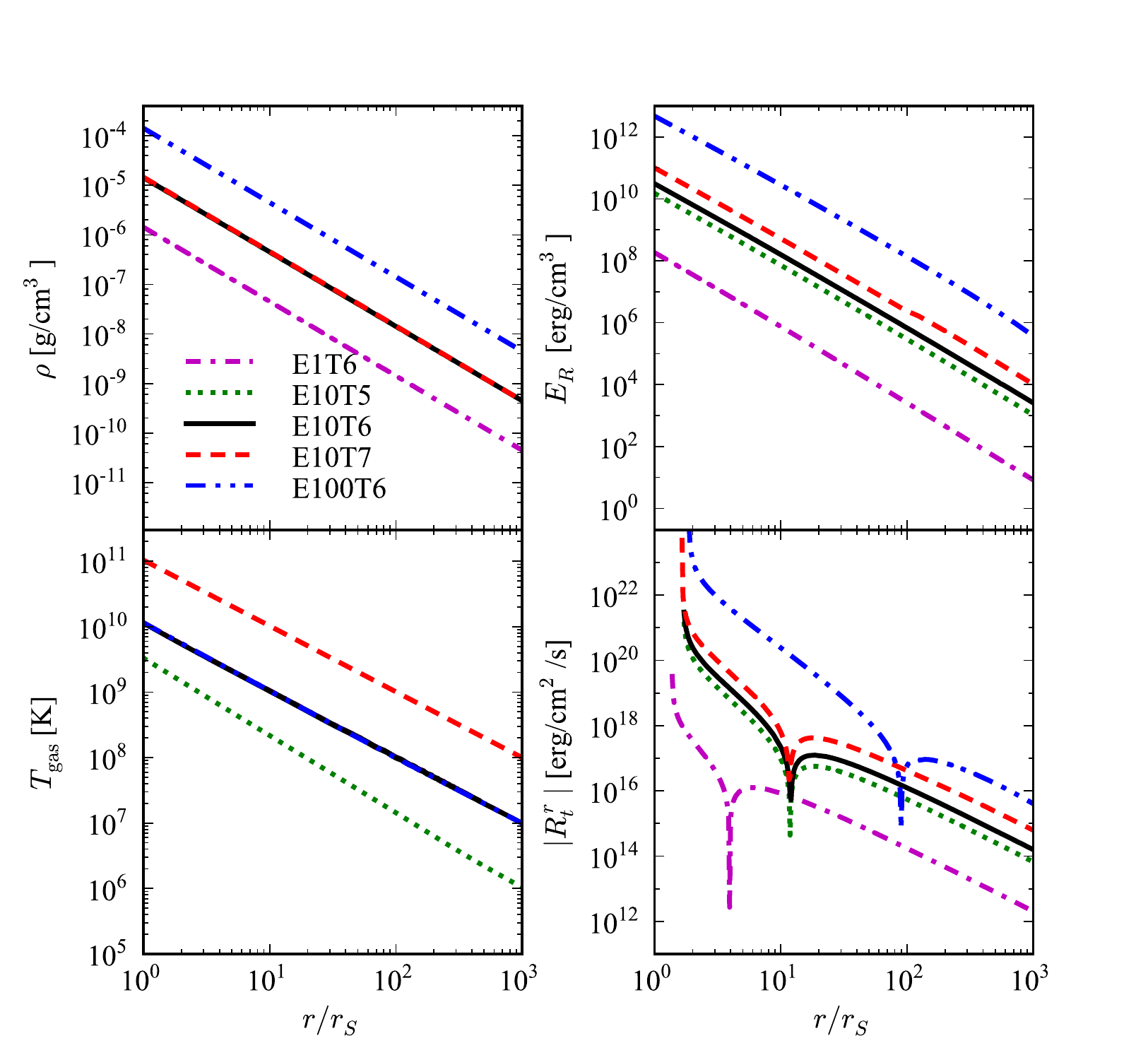}
\caption{Profiles of $\rho$, $T_\mathrm{gas}$, $E_R$, and $|R^r_t|$ for five different combinations of $\dot{m}$ and $T_o$ in the Bondi inflow problem.  The sharp dips in $|R^r_t|$ mark the photon trapping radius for each case.
\label{fig:bondi}}
\end{center}
\end{figure}

Table \ref{tab:bondi} summarizes the key simulation parameters for this section.  Each simulation is run to $t=10^4 M = 0.15$ s, long enough for the radiation energy density, $E_R$, and radiation energy flux, $R^r_t$, to achieve steady-state profiles out beyond $r=10^3 r_S$.  Profiles of $\rho$, $T_\mathrm{gas}$, $E_R$, and $|R^r_t|$ are shown in Figure \ref{fig:bondi} for the five cases we consider.  These profiles are very similar to the comparable cases in \citet{fragile12}.  The sharp dips in the profiles of $|R^r_t|$ in the lower-right panel indicate the photon trapping radius for each flow.  Inside this radius, the net radiation energy flux is negative (toward the black hole), whereas outside this radius, it is positive.

\begin{deluxetable}{cccccccc}
\tabletypesize{\scriptsize}
\tablecaption{Radiative Bondi Simulations \label{tab:bondi}}
\tablewidth{0pt}
\tablehead{
\colhead{Simulation} & \colhead{$\dot{m}$} & \colhead{$T_o$ (K)} & $l$
}
\startdata
E1T6 & 1 & $10^6$ & $5.33 \times 10^{-8}$ \\
E10T5 & 10 & $10^5$ & $1.75 \times 10^{-6}$ \\
E10T6 & 10 & $10^6$ & $4.13 \times 10^{-6}$ \\
E10T7 & 10 & $10^7$ & $1.63 \times 10^{-5}$ \\
E100T6 & 100 & $10^6$ & $1.06 \times 10^{-4}$
\enddata
\end{deluxetable}

An important point about the profiles in Figure \ref{fig:bondi} is that none of them show the dramatic oscillations that were seen in some cases in \citet{fragile12}.  Those oscillations were symptomatic of the instability of our previous method, especially in optically thin regions; their absence here is another indication that our current method is a significant improvement.  Another such indication is that, although these simulations ran to a similar end time as in our previous paper, they did so in about an order of magnitude fewer cycles.  This is a direct result of the larger timestep we are able to take when treating the source term implicitly.

The most interesting diagnostic to consider for these radiative Bondi flows is the emitted luminosity.  In the current work, this can most easily be recovered from the radiation energy flux $R^r_t$.  Specifically, 
\begin{equation}
L = -\int_S \sqrt{-g} R^r_t dA_r ~,
\end{equation}
where $dA_r$ is the surface area element normal to the radial direction.  We report the resulting luminosity, in units of the Eddington luminosity $l = L/L_\mathrm{Edd}$, where $L_\mathrm{Edd} = \dot{M}_\mathrm{Edd} c^2 = 4 \pi G M c \sigma_T/m_p$, in Table \ref{tab:bondi}.

\section{Two-Dimensional, Quasi-Spherical Inflow with Radiation}
\label{sec:proga}

We now consider a new application of our general relativistic radiation hydrodynamics method to the problem of quasi-spherical accretion onto a black hole.  The flow is quasi-spherical in the sense that we start with outer boundary conditions similar to the spherically-symmetric Bondi inflow problem discussed in the previous section, the only difference being that a small amount of angular momentum is added to the gas, thus breaking the symmetry.  In practice, we actually start from a two-dimensional version of the Bondi inflow problem with no angular momentum.  We run this for a time of $10^4 M = 0.15$ s to allow the radiation to reach an equilibrium before introducing angular momentum of the form \citep{proga03a}:
\begin{equation}
\ell = \ell_0 (1 - \vert \cos \theta \vert )~,
\end{equation}
where $l_0=2 l_\text{ms}$ and $l_\text{ms}$ is the specific angular momentum of a test particle orbiting in the equatorial plane at the innermost stable circular orbit (or ISCO).  We then run the simulation for an additional $10^4 M$ after the introduction of the angular momentum, keeping the new outer boundary conditions constant.

For these simulations we use a spherical-polar $(r,\theta)$ grid, still assuming symmetry about the rotational axis (three-dimensional simulations will be considered in future work). The radial range extends from $r_i=0.95r_\mathrm{S}$ to $r_o = 1000r_\mathrm{S}$, and the angular range covers $0 \le \theta \le \pi$. Our grid is discretized into $384 \times 192$ zones. We use the same logarithmic radial coordinate as in the Bondi problem and a uniform grid in the angular direction.  

\begin{figure}
\begin{center}
\includegraphics[width=0.49\columnwidth]{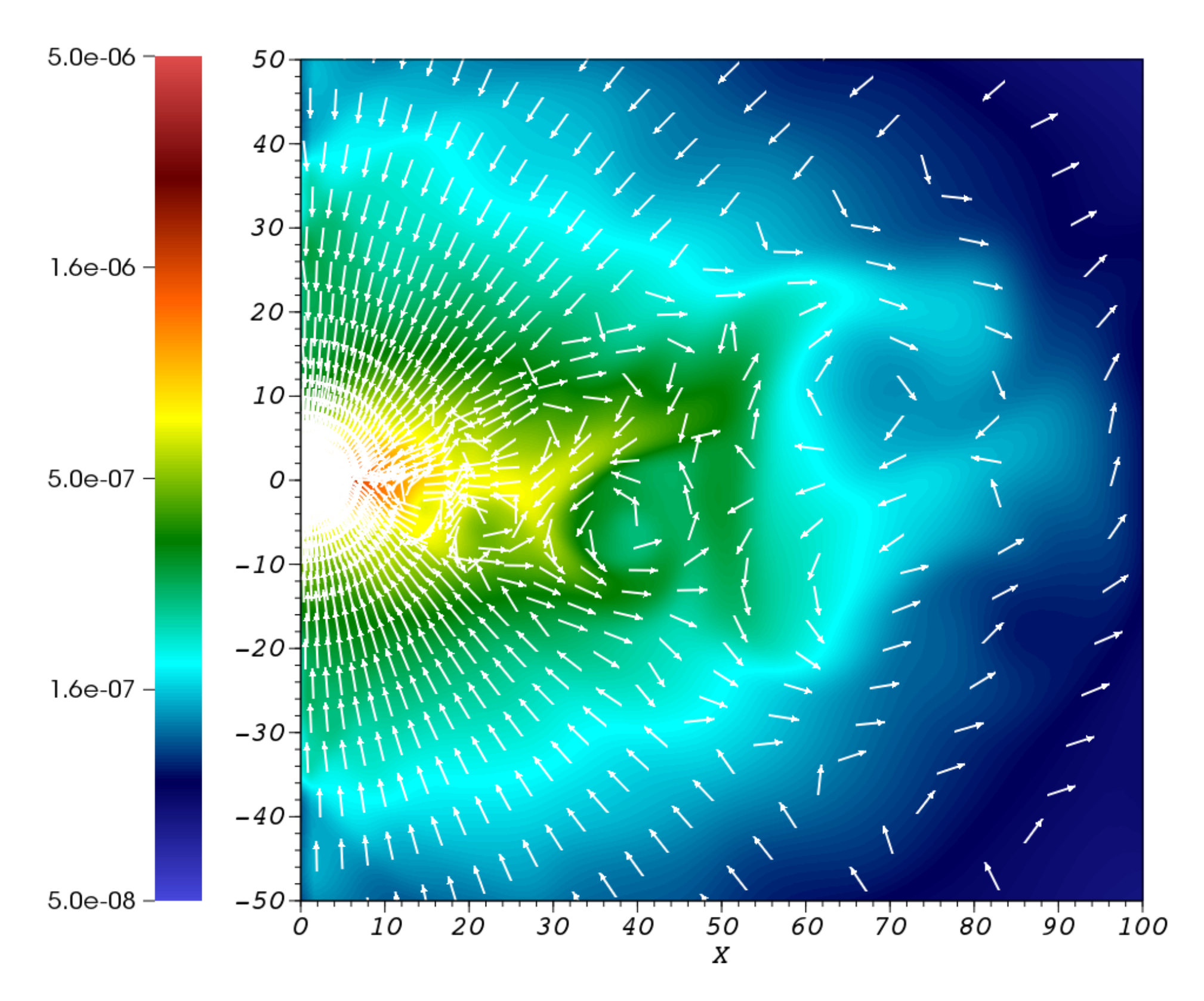} 
\includegraphics[width=0.49\columnwidth]{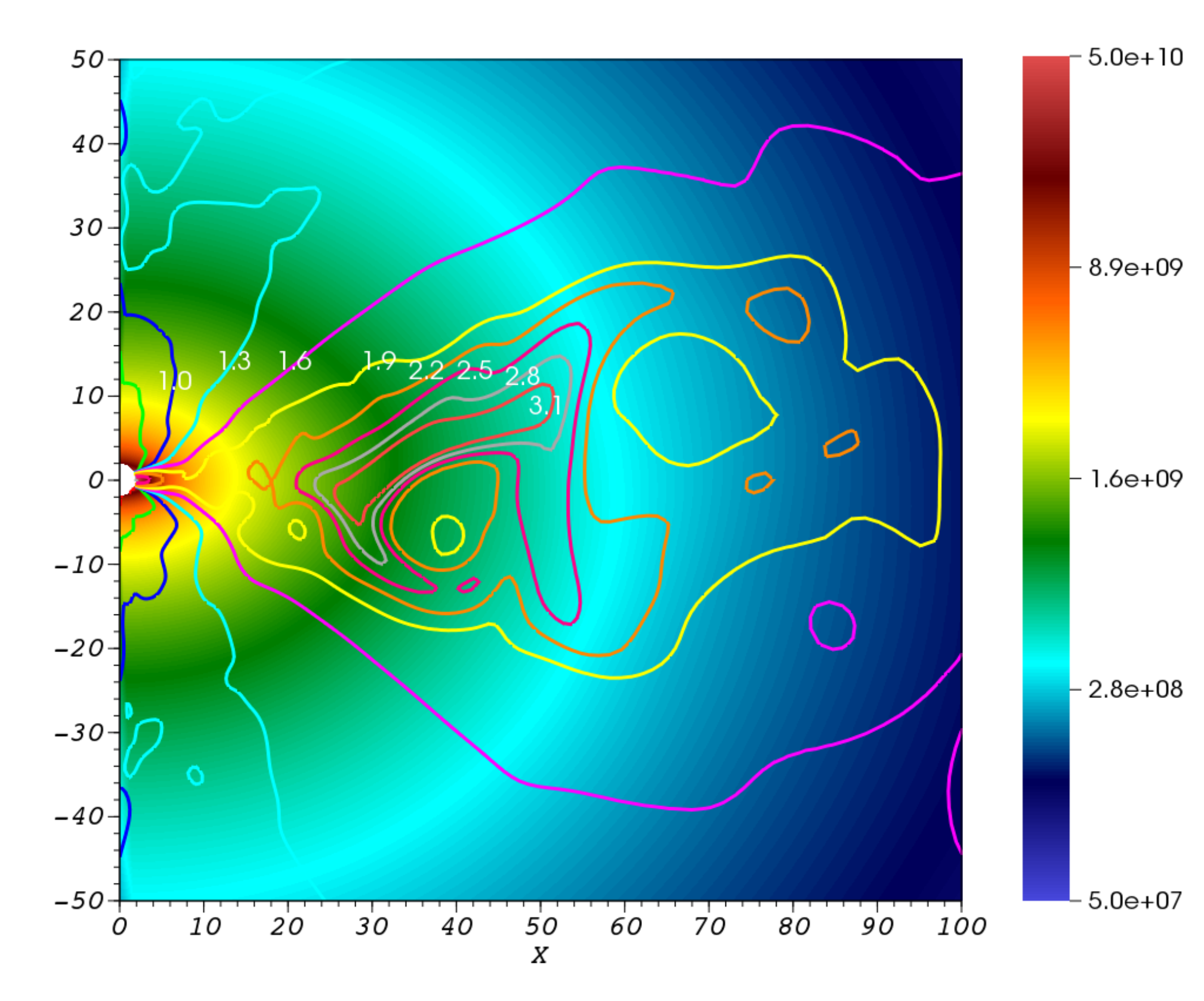} 
\caption{{\em Left panel}: Pseudocolor of $\rho$ (in units of g cm$^{-3}$), with vectors representing the local poloidal fluid velocity direction, at $t=16500 M = 0.24$ s for the two-dimensional quasi-spherical inflow problem.  {\em Right panel}: Pseudocolor of $E_R$ (in units of erg cm$^{-3}$), with contours representing the optical depth, $\tau$ (with 10 contour levels from 0.4 close to the poles to 3.1 near the equatorial plane at $r=30M$).
\label{fig:proga}}
\end{center}
\end{figure}

\begin{figure}
\begin{center}
\includegraphics[width=0.49\columnwidth]{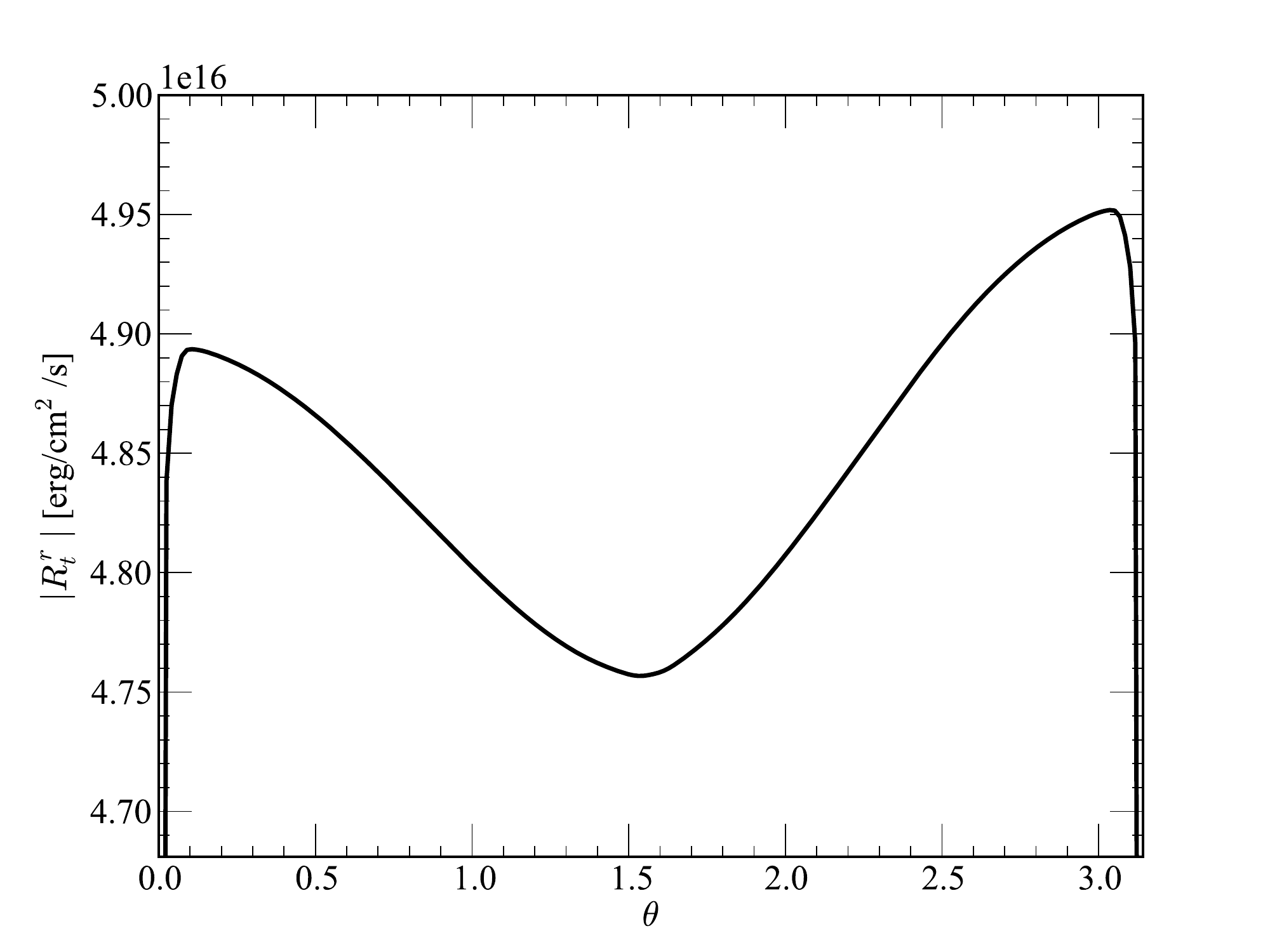} 
\caption{Radiation energy flux, $|R^r_t|$, as a function of angle at $r \approx 1000 r_S$ for the two-dimensional quasi-spherical inflow problem, time averaged over the duration of the simulation.  As usual, the angle $\theta$ runs from 0 at the ``north'' pole to $\pi/2$ at the midplane to $\pi$ at the ``south'' pole.  We see that the radiation flux is highest along the poles, as expected.
\label{fig:flux}}
\end{center}
\end{figure}

Along with breaking symmetry, the introduction of angular momentum provides some centrifugal support to the gas.  Conservation of angular momentum allows this support to become more significant as the gas is transported to smaller radii, leading to the formation of a thick, disk-like structure close to the black hole (see Figure \ref{fig:proga}, {\em left} panel).  The spherical symmetry of the radiation field is also broken (see Figure \ref{fig:proga}, {\em right} panel).  Figure \ref{fig:proga} also shows that the optical depth of the gas varies with latitude, being lower along the symmetry axis and higher near the midplane at a given radius.  Here we approximate the optical depth as $\tau \simeq \rho (\kappa^\mathrm{a} + \kappa^\mathrm{s}) r$.  Figure \ref{fig:flux} shows that this latitude dependence carries over to the radiative flux, $R^r_t$, even at large radii, with the flux varying by about 4\% from midplane to pole.  In more extreme cases, such a source could have significantly different inferred (isotropic) luminosities when viewed from different angles.  This kind of latitude-dependent luminosity could be important for understanding ultra-luminous X-ray sources (ULXs) \citep{komossa98, swartz04}, among other phenomena.

The total emitted luminosity at the outer grid boundary for this simulation is approximately $l = 1.5\times10^{-4}$ (where, again, $l$ is in units of the Eddington luminosity).  Despite the very high mass accretion rate, we again observe a very sub-Eddington luminosity, although not as low as for a one-dimensional Bondi inflow problem with the same mass accretion rate and temperature ($l_\mathrm{Bondi} = 2.3\times10^{-6}$), so the centrifugal support is allowing the gas to radiate more of its energy prior to being accreted into the black hole.  

\begin{figure}
\begin{center}
\includegraphics[width=0.49\columnwidth]{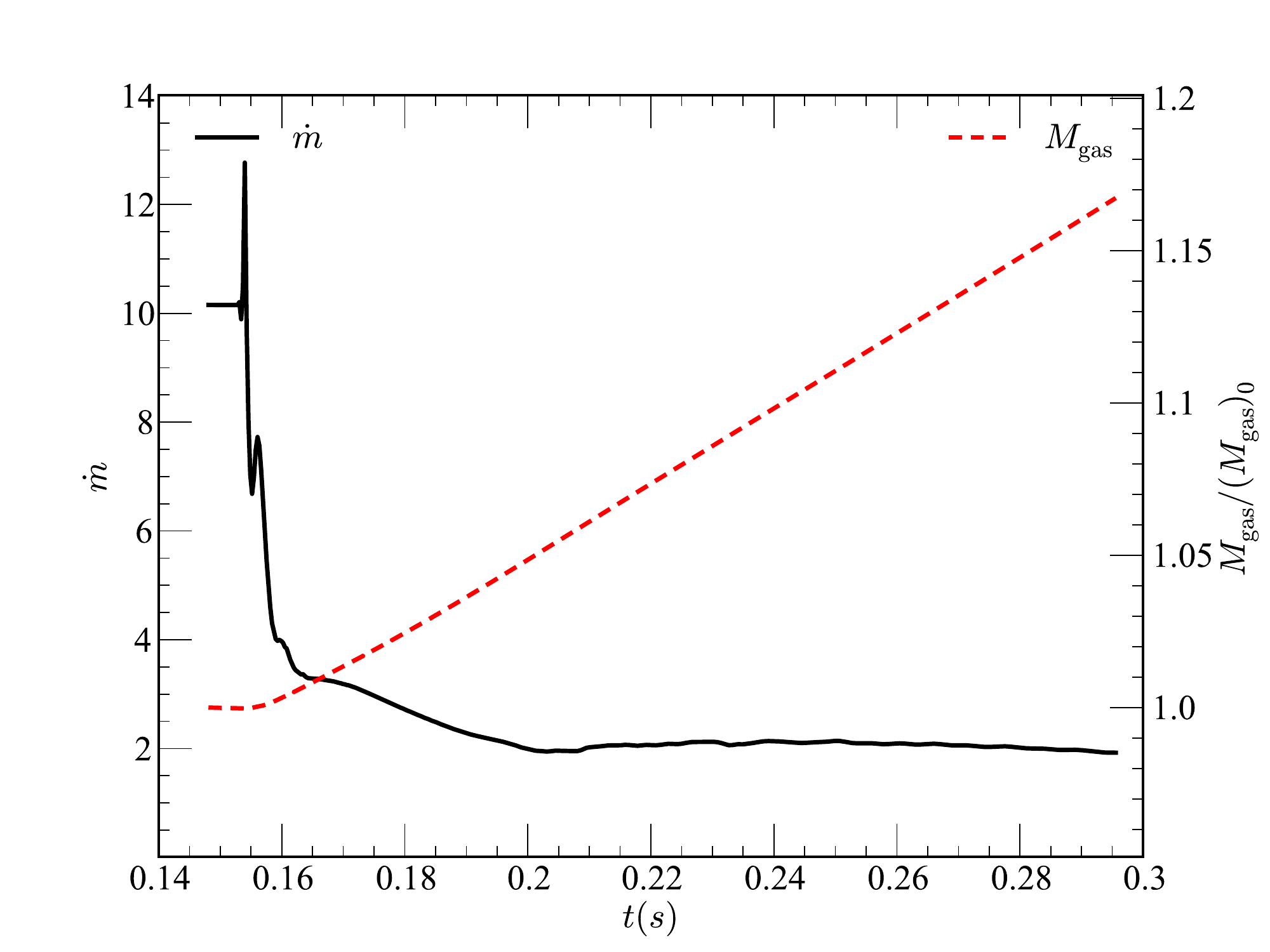} 
\caption{Mass flux through the event horizon (in units of $\dot{M}_\mathrm{Edd}$) as a function of time ({\em black solid} and {\em left axis}) and total mass on the grid ({\em red dashed} and {\em right axis}) for the two-dimensional quasi-spherical inflow problem.  Mass is fed from the outer boundary at a rate of $\dot{m}=10$.
\label{fig:mdot}}
\end{center}
\end{figure}

The centrifugal support and increased radiation pressure also means that not as much gas is actually able to reach the black hole event horizon as in the one-dimensional problem.  Figure \ref{fig:mdot} shows that the effective mass accretion rate onto the black hole is only about one fifth of the feeding rate at the outer boundary.  It also shows that the total gas mass on the grid continues to increase throughout the duration of the simulation, forming an ever larger disk.  One goal of our future work will be to study the long-term evolution of this and similar flows to determine what ultimately happens to this mass - is it subsequently accreted onto the hole or carried away in outflows? 

Finally, we note that there are interesting low density ``bubbles'' in the accretion disk, as seen in the {\em left} panel of Figure \ref{fig:proga}.  These bubbles are long-lived and slowly move outward in radius, possibly due to buoyant forces.  These features can not be gas pressure supported, as they exhibit lower gas pressure than their surroundings.  They also do not appear to be radiation pressure supported, since they do not show up in any of the radiation field plots (see, for example, the {\em right} panel of Figure \ref{fig:proga}).  Of course, because they are low density, they are also identifiable in the {\em right} panel of Figure \ref{fig:proga} as relatively low optical depth regions (remember, we calculate optical depth as $\tau \simeq \rho (\kappa^\mathrm{a} + \kappa^\mathrm{s}) r$).  It appears these features are associated with hydrodynamic eddies that form in the flow, as indicated by looking at the velocity vector field in the {\em left} panel of Figure \ref{fig:proga}.  Features such as these could conceivably play a role in enhancing the local radiative flux and overall radiative efficiency of the flow, although this effect does not appear to be significant in this case.  This could be because the density contrasts in this particular case are not very large (roughly a factor of 2).  It will be interesting to see how these features behave at higher luminosities and in the presence of magnetic fields, topics that will be explored in future work.

\section{Conclusion}
\label{sec:conclusion}

In this paper we have presented a new method for solving the equations of general relativistic radiation hydrodynamics.  Our testing has demonstrated two critical improvements over our previous method:  The first is the introduction of a more general closure scheme for the radiation equations, in our case, the so-called $\bf{M}_1$ closure.  As shown with the cloud shadow test (Section \ref{sec:cloud}), this scheme corrects some of the unphysical flaws found with the Eddington approximation used previously.  Furthermore, it allows us to cover a much broader range of optical depths, as demonstrated by the Bondi inflow test (Section \ref{sec:bondi}).  The second is the implementation of a hybrid explicit-implicit (or semi-implicit) evolution scheme.  The Bondi inflow test further demonstrated that our new method is stable over a much broader range of parameters than our previous one.  We are now able to consider temperatures and mass accretion rates that are orders of magnitude higher and lower than were possible in our earlier study, and often with significantly larger timesteps and reduced computational expense.  

Because of these improvements, we are now in a position to study entire classes of problems that were not accessible to our previous method, such as the cloud shadow test, and of more physical interest, multi-dimensional, super-Eddington accretion onto black holes.  An example of this latter class of problems was presented in Section \ref{sec:proga}, in the form of a quasi-spherical accretion flow onto a black hole, with a mass accretion rate ten times the Eddington value.  The angular momentum supplied to the gas in that case provided enough centrifugal support for a disk-like structure to form.  However, the radiative efficiency remained extremely low ($1.5\times10^{-4}$ in units of the Eddington luminosity).  As in the pure Bondi inflow case, most of the dissipated energy was carried into the black hole.  In contrast, more nearly Keplerian disks have been shown to exhibit (inferred isotropic) luminosities as high as $\sim 20 L_\mathrm{Edd}$ \citep{ohsuga11,sadowski14,mckinney14}.  Further simulations will help bridge the gap between our low angular momentum case and those higher angular momentum ones.  We can now also begin a systematic study of the parameter space associated with super-Eddington accretion.  There are many open theoretical questions to be addressed by such a study.  There is also a rich phenomenology of observed behavior in black hole systems accreting near the Eddington limit that have yet to be seen in simulations, providing another avenue for exploration.

\acknowledgements
We thank Aleksander S{\c a}dowski, Ken Ohsuga, and Eirik Endeve for their helpful feedback and discussions.  This work was supported in part by a High-Performance Computing grant from Oak Ridge Associated Universities/Oak Ridge National Laboratory and by the National Science Foundation under grants NSF AST-1211230 and NSF PHY11-25915.  AO gratefully acknowledges funding support from the College of Charleston office of Undergraduate Research and Creative Activities through grant SU2013-036.  The work by PA was performed under the auspices of the U.S. Department of Energy by Lawrence Livermore National Laboratory under Contract DE-AC52-AC52-07NA27344.  This research used resources of the Oak Ridge Leadership Computing Facility, located in the National Center for Computational Sciences at Oak Ridge National Laboratory, which is supported by the Office of Science of the Department of Energy under Contract DE-AC05-00OR22725.

\appendix

\section{1st Order Taylor Expansion Terms}
\label{sec:derivs}

As described in Section \ref{sec:source}, the Jacobian matrix, ${\bf A}$, can either be calculated analytically or numerically.  Although more tedious to code, we have found that the analytic method is consistently faster on all our tests, making it perhaps worth the extra effort.  To aid those who might wish to code the analytic solution, we record all the pertinent partial derivatives for equation (\ref{eqn:A}) here, ordered by conserved field.

Mass density:
\begin{align*}
  \frac{\partial D}{\partial \rho}            &= W \\
  \frac{\partial D}{\partial \widetilde{u}^i} &= \sqrt{-g} \rho \frac{\partial u^t}{\partial \widetilde{u}^i} \\
  \frac{\partial D}{\partial \epsilon}        &= \frac{\partial D}{\partial E_R} = \frac{\partial D}{\partial \widetilde{u}_R^i} = 0
\end{align*}

Fluid energy:
\begin{align*}
\frac{\partial {\cal E}}{\partial \rho}            &= -\sqrt{-g} \left[(1 + \epsilon) u^t u_t + (u^t u_t + 1) \frac{\partial  P_\mathrm{gas}}{\partial  \rho} \right] \\
\frac{\partial {\cal E}}{\partial  \epsilon}       &= -\sqrt{-g} \left[\rho u^t u_t + (u^t u_t + 1) \frac{\partial P_\mathrm{gas}}{\partial \epsilon} \right] \\
\frac{\partial {\cal E}}{\partial \widetilde{u}^i} &= -\sqrt{-g} (\rho h + 2 P_{mag}) 
                                                       \left(u_t \frac{\partial u^t}{\partial \widetilde{u}^i} + 
                                                             u^t \frac{\partial u_t}{\partial \widetilde{u}^i}\right) \\
\frac{\partial {\cal E}}{dE_R}                     &= \frac{\partial {\cal E}}{d\widetilde{u}_R^i} = 0
\end{align*}

Fluid momentum:
\begin{align*}
\frac{\partial{\cal S}_j}{\partial\rho}            &= \sqrt{-g} u^t u_j \left(1 + \epsilon + \frac{\partial P_\mathrm{gas}}{\partial \rho} \right) \\
\frac{\partial{\cal S}_j}{\partial\epsilon}        &= \sqrt{-g} u^t u_j \left(\rho + \frac{\partial P_\mathrm{gas}}{ \partial\epsilon} \right) \\
\frac{\partial{\cal S}_j}{\partial\widetilde{u}^i} &= \sqrt{-g} (\rho h + 2 P_{mag}) 
                                                      \left( u_j \frac{\partial u^t}{\partial \widetilde{u}^i} + 
                                                             u^t \frac{\partial u_j}{\partial \widetilde{u}^i} \right) \\
\frac{\partial{\cal S}_j}{\partial E_R}            &= \frac{\partial{\cal S}_j}{\partial\widetilde{u}_R^i} = 0
 \end{align*}

Radiation energy:
\begin{align*}
\frac{\partial{\cal R}}{\partial E_R}       &= \sqrt{-g} \left(\frac{4}{3}  u^t_R (u_R)_t + \frac{1}{3} \right) \\
\frac{\partial{\cal R}}{d\widetilde{u}^i_R} &= \sqrt{-g} \frac{4}{3} E_R 
                                               \left( (u_R)_t \frac{\partial u^t_R}{\partial \widetilde{u}^i_R} + 
                                                      u^t_R \frac{\partial(u_R)_t}{\partial \widetilde{u}^i_R} \right) \\
\frac{\partial{\cal R}}{\partial \rho}      &= \frac{\partial{\cal R}}{\partial \epsilon} 
                                             = \frac{\partial{\cal R}}{\partial \widetilde{u}^i} = 0
\end{align*}

Radiation momentum:
\begin{align*}
\frac{\partial{\cal R}_j}{\partial E_R}              &= \sqrt{-g} \frac{4}{3} u^t_R (u_R)_j  \\
\frac{\partial{\cal R}_j}{\partial\widetilde{u}^i_R} &= \sqrt{-g} \frac{4}{3} E_R 
                                                        \left( (u_R)_j \frac{\partial u^t_R}{\partial \widetilde{u}^i_R} + 
                                                               u^t_R \frac{\partial(u_R)_j}{\partial \widetilde{u}^i_R}\right) \\
\frac{\partial{\cal R}_j}{\partial\rho}              &= \frac{\partial{\cal R}_j}{\partial\epsilon} 
                                                      = \frac{\partial{\cal R}_j}{\partial \widetilde{u}^i} = 0
\end{align*}

Also appearing in the Jacobian are the following gradients of the radiation 4-force density:
\begin{align*}
\frac{\partial G_\mu}{\partial\rho} =& - \left(\kappa^\mathrm{a} + \kappa^\mathrm{s}\right) R_{\mu \nu} u^{\nu} - \left(\kappa^\mathrm{s} R_{\alpha \beta} u^\alpha u^\beta + \kappa^\mathrm{a} a_R T^4_\mathrm{gas}\right) u_\mu - \rho\left(R_{\mu \nu}u^\nu + a_R T^4_\mathrm{gas} u_\mu\right) \frac{\partial\kappa^\mathrm{a}}{\partial\rho} \\
 & - \rho \left(R_{\mu \nu}u^\nu + R_{\alpha \beta} u^\alpha u^\beta u_\mu\right)\frac{\partial\kappa^\mathrm{s}}{\partial\rho} - 4 \rho \kappa^\mathrm{a} a_R T^3_\mathrm{gas} u_\mu \frac{\partial T_\mathrm{gas} }{\partial\rho} \\
\frac{\partial G_{\mu} }{\partial\epsilon} =& -\rho \left(R_{\mu \nu}u^\nu + a_R T^4_\mathrm{gas} u_\mu\right) \frac{\partial\kappa^\mathrm{a}}{\partial\epsilon} - \rho \left(R_{\mu \nu} u^\nu + R_{\alpha\beta} u^\alpha u^\beta u_\mu\right)\frac{\partial\kappa^\mathrm{s}}{\partial\epsilon} - 4 \rho \kappa^\mathrm{a} a_R T^3_\mathrm{gas} u_\mu \frac{\partial T_\mathrm{gas}}{\partial\epsilon} \\
\frac{\partial G_\mu}{\partial\widetilde{u}^i} =& -\rho(\kappa^\mathrm{a} + \kappa^\mathrm{s}) R_{\mu \nu} \frac{\partial u^\nu}{\partial \widetilde{u}^i} - \rho \left(\kappa^\mathrm{s} R_{\alpha \beta} u^\alpha u^\beta + \kappa^\mathrm{a} a_R T^4_\mathrm{gas}\right) g_{\mu \nu} \frac{\partial u^\nu}{\partial \widetilde{u}^i} - \rho \kappa^\mathrm{s} u_\mu R_{\alpha \beta} \left( u^\alpha \frac{\partial u^\beta}{\partial \widetilde{u}^i} + u^\beta \frac{\partial u^\alpha}{\partial \widetilde{u}^i}\right)\\
\frac{\partial G_\mu}{\partial E_R} =& -\rho (\kappa^\mathrm{a} + \kappa^\mathrm{s}) u^\nu \frac{\partial R_{\mu\nu}}{\partial E_R} - \rho \kappa^\mathrm{s} u_\mu u^\alpha u^\beta \frac{\partial R_{\alpha \beta}}{\partial E_R} \\
\frac{\partial G_\mu}{\partial \widetilde{u}^i_R} =& -\rho (\kappa^\mathrm{a} + \kappa^\mathrm{s}) u^\nu \frac{\partial R_{\mu\nu}}{\partial \widetilde{u}^i_R} - \rho \kappa^\mathrm{s} u_\mu u^\alpha u^\beta \frac{\partial R_{\alpha \beta}}{\partial \widetilde{u}^i_R} \\
\end{align*}

Finally, the following partial derivatives are needed to evaluate the above expressions:
\begin{align*}
\frac{\partial u^t}{\partial \widetilde{u}^i}     &= \frac{1}{\gamma \alpha} g_{ij} \widetilde{u}^j \\
\frac{\partial u^j}{\partial \widetilde{u}^i}     &= \delta^j_i + \frac{g^{tj}}{g^{tt}} \frac{\partial u^t}{\partial \widetilde{u}^i} \\
\frac{\partial u^t_R}{\partial \widetilde{u}^i_R} &= \frac{1}{\gamma \alpha} g_{ij} \widetilde{u}^j_R \\
\frac{\partial u^j_R}{\partial \widetilde{u}^i_R} &= \delta^j_i + \frac{g^{tj}}{g^{tt}} \frac{\partial u^t_R}{\partial \widetilde{u}^i_R} \\
\frac{\partial R_{\alpha \beta}}{\partial E_R}    &= \frac{4}{3} (u_R)_\alpha (u_R)_\beta + \frac{1}{3} g_{\alpha \beta} \\
\frac{\partial R_{\alpha\beta}}{\partial \widetilde{u}^i_R} &= \frac{4}{3} E_R 
                                                               \left[ (u_R)_\alpha \frac{\partial (u_R)_\beta}{\partial \widetilde{u}^i_R} + 
                                                                      (u_r)_\beta \frac{\partial (u_R)_\alpha}{\partial \widetilde{u}^i_R}\right]
\end{align*}

\begin{thebibliography}{24}
\expandafter\ifx\csname natexlab\endcsname\relax\def\natexlab#1{#1}\fi

\bibitem[{{Anninos} {et~al.}(2005){Anninos}, {Fragile}, \&
  {Salmonson}}]{anninos05}
{Anninos}, P., {Fragile}, P.~C., \& {Salmonson}, J.~D. 2005, \apj, 635, 723

\bibitem[{{Dibi} {et~al.}(2012){Dibi}, {Drappeau}, {Fragile}, {Markoff}, \&
  {Dexter}}]{dibi12}
{Dibi}, S., {Drappeau}, S., {Fragile}, P.~C., {Markoff}, S., \& {Dexter}, J.
  2012, \mnras, 426, 1928

\bibitem[{{Drappeau} {et~al.}(2013){Drappeau}, {Dibi}, {Dexter}, {Markoff}, \&
  {Fragile}}]{drappeau13}
{Drappeau}, S., {Dibi}, S., {Dexter}, J., {Markoff}, S., \& {Fragile}, P.~C.
  2013, \mnras, 431, 2872

\bibitem[{{Farris} {et~al.}(2008){Farris}, {Li}, {Liu}, \&
  {Shapiro}}]{farris08}
{Farris}, B.~D., {Li}, T.~K., {Liu}, Y.~T., \& {Shapiro}, S.~L. 2008, \prd, 78,
  024023

\bibitem[{{Fragile} {et~al.}(2012){Fragile}, {Gillespie}, {Monahan},
  {Rodriguez}, \& {Anninos}}]{fragile12}
{Fragile}, P.~C., {Gillespie}, A., {Monahan}, T., {Rodriguez}, M., \&
  {Anninos}, P. 2012, \apjs, 201, 9

\bibitem[{{Fragile} \& {Meier}(2009)}]{fragile09}
{Fragile}, P.~C., \& {Meier}, D.~L. 2009, \apj, 693, 771

\bibitem[{{Hayes} \& {Norman}(2003)}]{hayes03}
{Hayes}, J.~C., \& {Norman}, M.~L. 2003, \apjs, 147, 197

\bibitem[{{Jiang} {et~al.}(2012){Jiang}, {Stone}, \& {Davis}}]{jiang12}
{Jiang}, Y.-F., {Stone}, J.~M., \& {Davis}, S.~W. 2012, \apjs, 199, 14

\bibitem[{{Komossa} \& {Schulz}(1998)}]{komossa98}
{Komossa}, S., \& {Schulz}, H. 1998, \aap, 339, 345

\bibitem[{{Lentz} {et~al.}(2012){Lentz}, {Mezzacappa}, {Bronson Messer},
  {Liebend{\"o}rfer}, {Hix}, \& {Bruenn}}]{lentz12}
{Lentz}, E.~J., {Mezzacappa}, A., {Bronson Messer}, O.~E., {Liebend{\"o}rfer},
  M., {Hix}, W.~R., \& {Bruenn}, S.~W. 2012, \apj, 747, 73

\bibitem[{{Levermore}(1984)}]{levermore84}
{Levermore}, C.~D. 1984, JQSRT, 31, 149

\bibitem[{{McKinney} {et~al.}(2014){McKinney}, {Tchekhovskoy}, {Sadowski}, \&
  {Narayan}}]{mckinney14}
{McKinney}, J.~C., {Tchekhovskoy}, A., {Sadowski}, A., \& {Narayan}, R. 2014,
  \mnras, 441, 3177

\bibitem[{{M{\"u}ller} {et~al.}(2010){M{\"u}ller}, {Janka}, \&
  {Dimmelmeier}}]{muller10}
{M{\"u}ller}, B., {Janka}, H.-T., \& {Dimmelmeier}, H. 2010, \apjs, 189, 104

\bibitem[{{Noble} {et~al.}(2006){Noble}, {Gammie}, {McKinney}, \& {Del
  Zanna}}]{noble06}
{Noble}, S.~C., {Gammie}, C.~F., {McKinney}, J.~C., \& {Del Zanna}, L. 2006,
  \apj, 641, 626

\bibitem[{{Ohsuga} \& {Mineshige}(2011)}]{ohsuga11}
{Ohsuga}, K., \& {Mineshige}, S. 2011, \apj, 736, 2

\bibitem[{{Proga} \& {Begelman}(2003)}]{proga03a}
{Proga}, D., \& {Begelman}, M.~C. 2003, \apj, 582, 69

\bibitem[{{Roedig} {et~al.}(2012){Roedig}, {Zanotti}, \& {Alic}}]{roedig12}
{Roedig}, C., {Zanotti}, O., \& {Alic}, D. 2012, \mnras, 426, 1613

\bibitem[{{Rybicki} \& {Lightman}(1986)}]{rybicki86}
{Rybicki}, G.~B., \& {Lightman}, A.~P. 1986, {Radiative Processes in
  Astrophysics}

\bibitem[{{S{\c a}dowski} {et~al.}(2014){S{\c a}dowski}, {Narayan}, {McKinney},
  \& {Tchekhovskoy}}]{sadowski14}
{S{\c a}dowski}, A., {Narayan}, R., {McKinney}, J.~C., \& {Tchekhovskoy}, A.
  2014, \mnras, 439, 503

\bibitem[{{S{\c a}dowski} {et~al.}(2013){S{\c a}dowski}, {Narayan},
  {Tchekhovskoy}, \& {Zhu}}]{sadowski13}
{S{\c a}dowski}, A., {Narayan}, R., {Tchekhovskoy}, A., \& {Zhu}, Y. 2013,
  \mnras, 429, 3533

\bibitem[{{Shibata} {et~al.}(2011){Shibata}, {Kiuchi}, {Sekiguchi}, \&
  {Suwa}}]{shibata11}
{Shibata}, M., {Kiuchi}, K., {Sekiguchi}, Y., \& {Suwa}, Y. 2011, Progress of
  Theoretical Physics, 125, 1255

\bibitem[{{Swartz} {et~al.}(2004){Swartz}, {Ghosh}, {Tennant}, \&
  {Wu}}]{swartz04}
{Swartz}, D.~A., {Ghosh}, K.~K., {Tennant}, A.~F., \& {Wu}, K. 2004, \apjs,
  154, 519

\bibitem[{{Turner} \& {Stone}(2001)}]{turner01}
{Turner}, N.~J., \& {Stone}, J.~M. 2001, \apjs, 135, 95

\bibitem[{{Zanotti} {et~al.}(2011){Zanotti}, {Roedig}, {Rezzolla}, \& {Del
  Zanna}}]{zanotti11}
{Zanotti}, O., {Roedig}, C., {Rezzolla}, L., \& {Del Zanna}, L. 2011, \mnras,
  417, 2899

\end{thebibliography}

\end{document}